\pdfoutput=1
\documentclass[aps,amsmath,floats,floatfix,twocolumn,superscriptaddress,nofootinbib,showpacs,longbibliography]{revtex4-1}

\usepackage{packages}

\begin{document}

\title{Towards a few percent measurement of the Hubble constant with the current network of gravitational wave detectors without using electromagnetic information}

\newcommand{\ldit}{\affiliation{Laboratoire des 2 Infinis - Toulouse (L2IT-IN2P3), Université de Toulouse, CNRS, F-31062 Toulouse Cedex 9, France}}
\newcommand{\lpens}{\affiliation{Laboratoire de Physique de l’Ecole Normale Supérieure (LPENS), ENS, Université PSL, CNRS, 75005 Paris, France}}

\author{Tom Bertheas}\email{tom.bertheas@l2it.in2p3.fr}\ldit\lpens
\author{Vasco Gennari}\email{vasco.gennari@l2it.in2p3.fr}\ldit
\author{Nicola Tamanini}\ldit

\hypersetup{pdfauthor={Bertheas et al.}}

\date{\today}

\begin{abstract}

Gravitational waves (GWs) provide a novel and independent measurement of cosmological parameters, offering a promising avenue to address the Hubble tension alongside traditional electromagnetic observations.
In the absence of electromagnetic counterparts or complete host galaxy catalogs, current measurements rely on population-based methods that statistically combine black hole merger events.
Building on recent models that incorporate additional structure in the primary black hole mass distribution, using public data from the LIGO–Virgo–KAGRA~(LVK) collaboration third observing run (O3), we obtain a 30\% accuracy improvement on the measurement of the Hubble constant with respect to the result reported by LVK with the third GW transient catalog (GWTC-3).
Employing a realistic simulation that includes full Bayesian single-event inference, we present forecasts for the upcoming LVK observational runs, O4 and O5.
Using a three power-law mass model, we project a measurement of the Hubble constant with 20\% accuracy at O4 sensitivity, improving to 2.7\% accuracy at O5 sensitivity.
Our findings demonstrate the potential for gravitational waves to provide a substantial contribution to solving the Hubble tension within the next decade of observations.
\end{abstract}

\maketitle

\section{Introduction}
\label{sec:introduction}

In the last ten years since their first direct detection \cite{LIGOScientific:2016aoc}, gravitational waves (GWs) have firmly established a new observational window onto the Universe.
This plethora of new observations is yielding original insights into different physical and astronomical domains~\cite{LIGOScientific:2018mvr, LIGOScientific:2021usb, LIGOScientific:2020ibl, KAGRA:2021vkt, LIGOScientific:2020kqk, KAGRA:2021duu}.
GWs are in particular becoming a new instrument for cosmology, providing independent inference on the dynamics of the Universe uniquely complementary to standard electromagnetic (EM) probes~\cite{LIGOScientific:2017adf, LIGOScientific:2019zcs, LIGOScientific:2021aug}.
In fact, GWs naturally provide a direct measurement of the source's \textit{luminosity distance}, which, if combined with a \textit{redshift} estimate, can be directly used to measure cosmological parameters~\cite{Schutz:1986gp, Dalal:2006qt, Nissanke:2013fka}.

Obtaining complementary redshift information to a GW source is the main challenge of performing cosmological inference with GW signals, which per se do not directly deliver a redshift measurement.
Several approaches have so far been developed to provide redshift information.
The confident detection of an electromagnetic (EM) counterpart, allowing for a single galaxy to be associated to the GW signal (a so-called \textit{bright siren} event), provides exquisite cosmological constraints at low redshift~\cite{LIGOScientific:2017adf, LIGOScientific:2018hze, Palmese:2023beh}.
Around 50 of such bright sirens, could provide a 2\% measurement of the \textit{Hubble constant}, the parameter that determines the present rate of expansion of the Universe~\cite{Chen:2017rfc}.
Such a result would deliver a new, concrete insight into the so-called \textit{Hubble tension}~\cite{DiValentino:2021izs, Verde:2023lmm}, possibly hinting at new physics beyond the current standard cosmological model.
Unfortunately in the three GW observational campaigns completed so far by the LIGO-Virgo-KAGRA (LVK) collaboration~\cite{LIGOScientific:2018mvr, LIGOScientific:2021usb, LIGOScientific:2020ibl, KAGRA:2021vkt}, only one bright siren has been detected~\cite{LIGOScientific:2017ync}, and no new ones have so far been reported in the ongoing fourth observational run.

In the absence of an EM counterpart, cosmological inference with GWs can be performed accumulating information from multiple detections (generally referred to as \textit{dark sirens}) combined with archival EM data.
One such approach consists in using galaxy catalogs to statistically assign a host to GW events~\cite{Schutz:1986gp, Holz:2005df, DelPozzo:2011vcw, Gray:2023wgj}, another to cross correlate GW events to the matter distribution in the Universe~\cite{Namikawa:2015prh, Oguri:2016dgk, Mukherjee:2022afz}.
Combining multiple observations in this way require however to consistently infer the underlying events population, implying that a joint analysis of both cosmological and population parameters is required~\cite{Mastrogiovanni:2023emh, Gray:2023wgj}.
Remarkably, even in the absence of any EM information (direct counterparts or archival catalogs), a suitable modelling of the underlying GW population can still be exploited to infer cosmological parameters.
Such a method was initially proposed for binary neutron star mergers~\cite{Markovic:1993cr, Chernoff:1993th, Finn:1995wx, Taylor:2011fs}, and later extended to binary black holes (BBHs)~\cite{Farr:2019twy, Ezquiaga:2022zkx, Mastrogiovanni:2022hil, Mali:2024wpq}.
This approach is commonly referred to as \textit{spectral sirens}.

Given the lack of GW events with counterparts and being current all-sky galaxy catalogs limited by 
low completeness~\cite{LIGOScientific:2021aug}, spectral siren methods are fundamental to guarantee effective cosmological constraints with GWs from forthcoming LVK observations~\cite{KAGRA:2013rdx}.
The best constraints from spectral sirens based on GWTC-3 give a $\sim 50\%$ precision measurement on the Hubble constant, $H_0$, at $68\%$ confidence interval (C.I.), using 42 BBH events up to redshift $z \sim 0.8$~\cite{LIGOScientific:2021aug}.
These results are still far from the $\sim 15\%$ result obtained with the single bright siren detected so far~\cite{LIGOScientific:2017adf}\footnote{Depending on the astrophysical assumptions on peculiar velocities, neutron star spins and jet emission angle, the constraints can vary~\cite{Nicolaou:2019cip, LIGOScientific:2018hze, Hotokezaka:2018dfi, Palmese:2023beh, Mukherjee:2019qmm}.}.
The recent cosmological analysis by LVK of the newly released fourth version of the gravitational waves transient catalog (GWTC-4.0)~\cite{LIGOScientific:2025hdt, LIGOScientific:2025slb, LIGOScientific:2025yae} which includes events from the first part of the ongoing fourth observational run (O4a), yield $H_0 = 76.4^{+23.0}_{-18.1} ~\rm km \,s^{-1} \,Mpc^{-1}$ ($68\%$ C.I.) using the full population of detected GW sources (composed of both BBHs as well as binaries involving neutron stars) and $H_0 = 77.1^{+40.8}_{-26.3} ~\rm km \,s^{-1} \,Mpc^{-1}$ when restricting to BBH events, for a total of 142 and 137 events respectively~\cite{LIGOScientific:2025jau}. Those constraints bring down the accuracy of $H_0$ measurements to 27\% and 44\% respectively.
%

Spectral siren analyses adopt a \textit{parametric} approach to model the astrophysical distribution of BBH mergers, which requires specifying a functional form for the intrinsic distribution of GW parameters.
For the primary black hole (BH) mass, a combination of a power-law and a Gaussian (\plg) is  considered the simplest model able to capture the overall phenomenology of the BBH population reported in GWTC-3~\cite{LIGOScientific:2020kqk, KAGRA:2021duu}.
Nonetheless, several more agnostic, semi-parametric approaches have consistently shown the presence of additional structure, resolving, in addition to the fiducial \plg model, a sharp low-mass peak and a third overdensity at $\sim 20 M_{\odot}$~\cite{KAGRA:2021duu, Tiwari:2021yvr, Sadiq:2021fin, Farah:2023vsc, Toubiana:2023egi}.
These findings were recently confirmed in a fully parametric fashion, when Ref.~\cite{Gennari:2025nho} found that a primary mass model made of three power-laws (\camel) is able to reproduce such features, besides being preferred over the \plg by Bayesian evidence on GWTC-3 data.
Although recent GWTC-4.0 data now favors more intricate models than the fiducial \plg, LVK parametric analyses still struggle to resolve the additional structure in the BBH mass distribution described above, which is captured only by more flexible semi-parametric methods, which indeed hints at the presence of additional over-densities at $\sim 20 M_{\odot}$ and $\sim 60 M_{\odot}$~\cite{LIGOScientific:2025pvj, MaganaHernandez:2025cnu}. 

In this paper, we investigate the impact of multiple features in the BBH mass function on the estimation of cosmological parameters.
Analysing 50 events from the third LVK observing run (O3) with the \camel model, we report a measurement of $H_0 = 84^{+35}_{-25} ~\rm km \,s^{-1} \,Mpc^{-1}$ at 68\% C.I, hence a 35\% accuracy.
To our knowledge, this is the best constraint on the Hubble constant from GWs alone, i.e.~without using EM information, obtained from LVK O3 data.
In particular, it cuts down by a factor $\roughly 1.5$ the GWTC-3 constraints of $\roughly 50\%$ accuracy reported by the LVK collaboration using similar techniques but a \plg-like mass model\footnote{The data release~\cite{gwtc3:cosmo_release} associated to~\cite{LIGOScientific:2021aug} provides posterior samples from the \icarogw spectral sirens analysis for detection thresholds of $\rm SNR = 10/11/12$.
The corresponding constraints at 68\% CI are $H_0 = 79^{+38}_{-28} / 67^{+51}_{-35} / 67^{+41}_{-29} ~\rm km \,s^{-1} \,Mpc^{-1}$. In the main text, we compare our results with their $\rm SNR = 12$ analysis, and show the associated posterior in Fig.~\ref{fig:H0_posteriors_O3}. Note that the constraints reported in~\cite{LIGOScientific:2021aug}, namely $H_0 = 50^{+37}_{-30} ~\rm km \,s^{-1} \,Mpc^{-1}$, correspond to the \texttt{gwcosmo}~\cite{Gray:2019ksv, Gray:2021sew, Gray:2023wgj} spectral siren analysis while also inferring the dark energy equation of state parameter.}~\cite{LIGOScientific:2021aug}.
As discussed in Sec.~\ref{sec:summary} our results are comparable with the cosmological constraints recently reported by the LVK collaboration using GWTC-4.0 data~\cite{LIGOScientific:2025jau}.
A detailed comparison with these new O4a data requires extending our analyses to the population of binary neutron stars as well as to other BBH population models.
This requires a dedicated investigation which we leave for a future work.
%

We additionally provide forecasts for the upcoming LVK observing runs (a full 2 years O4 scenario, and an O5 scenario), to estimate our future ability to constrain $H_0$ with only GW information.
We predict a 20\% (40\%) measurement using a \camel (\plg) mass model for $\roughly 150$ observed events at O4 sensitivity, and a 2.7\% (18\%) measurement with $\roughly 1800$ events at O5 sensitivity (all at 68\% C.I.).
These results clearly show that a few percent measurement of the Hubble constant without EM information is within reach of O5, hinting at a possible solution of the Hubble tension with the current generation of GW detectors.

Unless explicitly specified, in what follows all results are reported at 68\% C.I.

\section{Real data constraints from O3}
\label{sec:O3_data}

\begin{figure}
    \hspace*{-0.3cm}\includegraphics[scale=0.6]{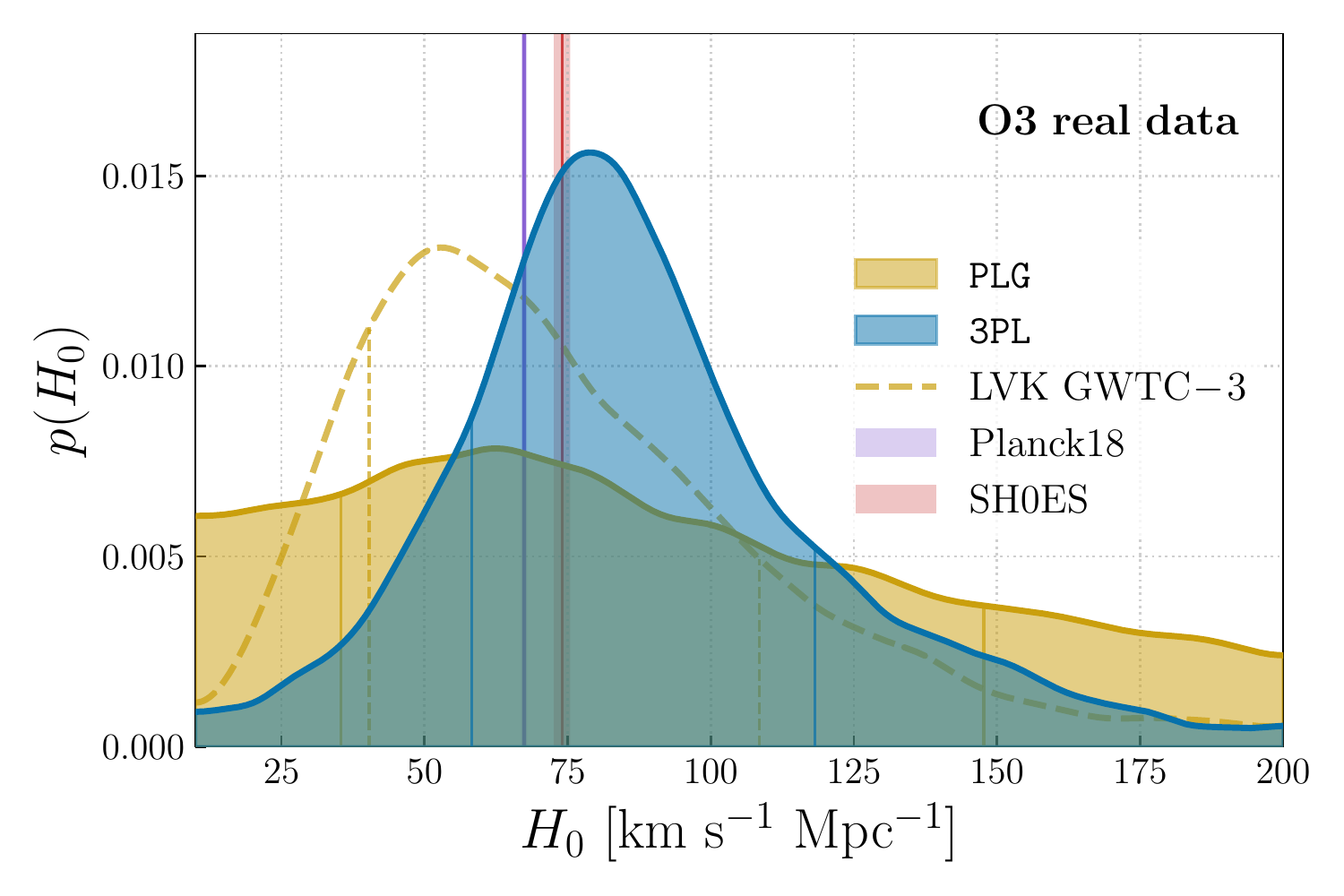}
    \caption{\justifying \footnotesize Marginalized posteriors on the Hubble constant from O3 real data analyses. Results using the \plg (\camel) model are shown in yellow (blue). The vertical lines in the reconstructed $H_0$ posterior distribution denote the $68\%$ C.I. The dashed posterior distribution shows \plg result from LVK~\cite{LIGOScientific:2021aug}. The purple and red shaded areas identify the 68\% C.I.~constraints on $H_0$ inferred from the CMB anisotropies by Planck~\cite{Planck:2018vyg} and in the local Universe by SH0ES~\cite{Riess:2021jrx}. (\textit{Colors available online})}
    \label{fig:H0_posteriors_O3}
\end{figure}

We employ \icarogw~\cite{Mastrogiovanni:2022hil, Mastrogiovanni:2023emh, Mastrogiovanni:2023zbw} to hierarchically combine events from the O3 observing run, simultaneously inferring the population parameters for the BBH distribution, the Hubble constant $H_0$ and the matter density at present day $\Omega_{m,0}$, assuming a flat $\Lambda$CDM cosmology.
Following Ref.~\cite{Gennari:2025nho}, we select 50 GW events with Inverse False Alarm Rate (IFAR) $> 4\, {\rm yr}$.
For the primary BH distribution, we employ the \plg and \camel mass models.
In Ref.~\cite{Gennari:2025nho}, the \camel model was found to be largely favoured on O3 data by Bayesian evidence over the \plg\footnote{In Ref.~\cite{Gennari:2025nho}, it was found that the \plg with wider priors than those used in LVK analyses, especially on the power-law index and the Gaussian width, provides comparable evidence to the \camel, but is not able to resolve the $\sim 35 M_{\odot}$ overdensity.}, with a Bayes factor of $\roughly 250$.
We adopt a power-law (truncated Gaussian) for the mass ratio distribution with the \plg (\camel), and a Madau-Dickinson~\cite{Madau:2014bja} for the rate evolution of the events.
The specific functional form of these parametrizations are described in Ref.~\cite{Gennari:2025nho}, to which we refer for further details about the methodology we employ. 
The priors used on the population parameters in O3 analyses are listed in Tab.~\ref{tab:models_priors} of App.~\ref{app:priors}.

\begin{figure*}[ht!]
    \centering
    \includegraphics[width=\textwidth]{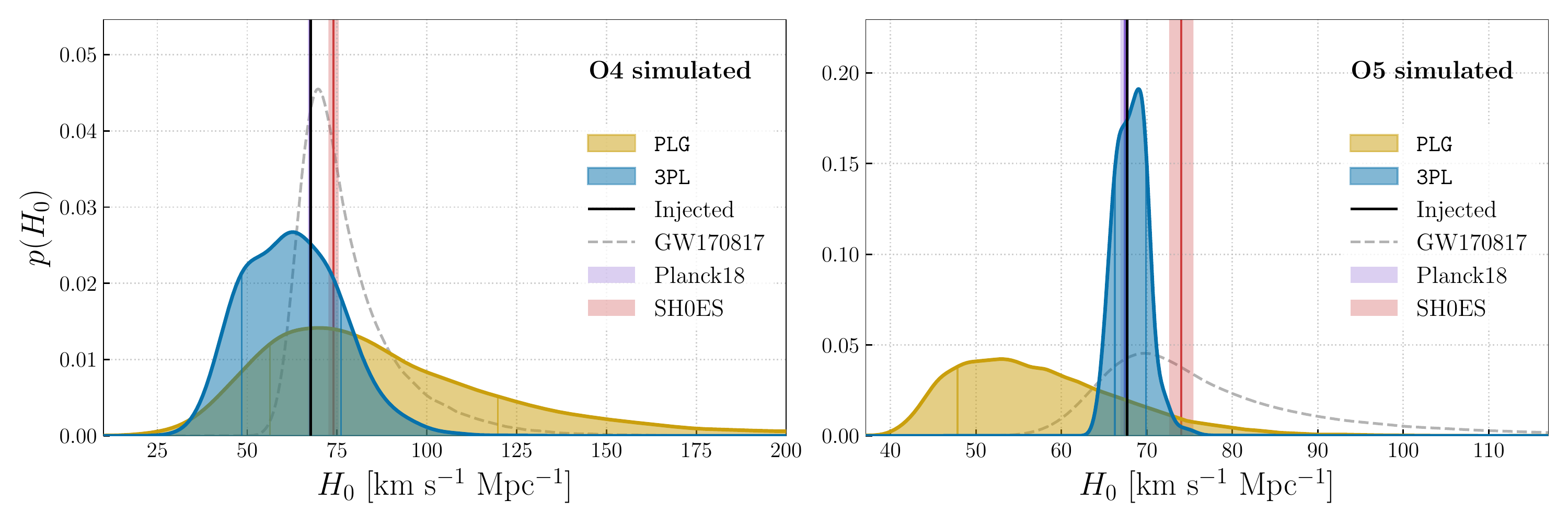}
    
    \caption{\justifying \footnotesize Marginalized posteriors on the Hubble constant from O4 (left) and O5 (right) simulations. Results using the \plg (\camel) model are shown in yellow (blue). The thin vertical lines in the reconstructed $H_0$ posterior denote the $68\%$ C.I. The purple and red shaded areas identify the $68\%$ C.I.~constraints on $H_0$ inferred by Planck~\cite{Planck:2018vyg} and SH0ES~\cite{Riess:2021jrx}, respectively. For comparison, the posterior distribution from GW170817~\cite{LIGOScientific:2017adf} is also reported in dashed grey. (\textit{Colors available online})}
    \label{fig:H0_posteriors}
\end{figure*}

Fig.~\ref{fig:H0_posteriors_O3} compares the marginal posterior distributions obtained on $H_0$ from our analysis, with the one reported by LVK results on GWTC-3~\cite{LIGOScientific:2021aug}.
For reference, the current estimates from the Planck~\cite{Planck:2018vyg} and SH0ES~\cite{Riess:2021jrx} collaborations are also shown.
With the \camel (\plg), we obtain an estimate of $H_0 = 84^{+35}_{-25} ~\rm km \,s^{-1} \,Mpc^{-1}$ ($H_0 = 81^{+67}_{-46} ~\rm km \,s^{-1} \,Mpc^{-1}$).
%
%
$H_0$ is better constrained by a factor $\roughly 1.5$ with our \camel analysis compared to the result reported by the LVK collaboration, which found a value of $H_0 = 67^{+41}_{-29} ~\rm km \,s^{-1} \,Mpc^{-1}$~\cite{LIGOScientific:2021aug} (our result reaches 35\% accuracy on the measurement of $H_0$, compared to LVK which achieved 52\% accuracy).
In addition to the improvement in accuracy, we observe a shift towards larger values of $H_0$ when using the \camel.
This behaviour is likely due to the \camel model ability to resolve the $10M_{\odot}$ mass peak more effectively than the \plg model, which tends to underestimate it~\cite{Gennari:2025nho}.
We also note that our \plg analysis produces broader posteriors compared to LVK.
This can be explained by the different datasets used and by the different priors adopted in our analysis, which are less stringent than those used by the LVK collaboration for the position of the Gaussian~\cite{Gennari:2025nho}.
We do not observe any significant improvement in the measurement of $\Omega_{m,0}$.

For completeness, we also mention the results obtained with other population models.
A model with a power-law and two Gaussian peaks~\cite{KAGRA:2021duu}, besides being disfavoured by evidence on GWTC-3 compared to the \plg~\cite{Gennari:2025nho}, provides similar constraints on $H_0$ to the \plg.
A \camel distribution with the features smoothed by a window function (see~\cite{Gennari:2025nho} for details about the corresponding parametrization) recovers comparable constraints than the \camel, $H_0 = 64^{+36}_{-29}~\rm km \,s^{-1} \,Mpc^{-1}$, but is penalized in evidence by the additional parameters. Note that this model is now the preferred model to describe the BBH population observed in GTWC-4.0 among those explored by LVK~\cite{LIGOScientific:2025pvj, LIGOScientific:2025jau}.

\section{Prospects for O4 and O5}
\label{sec:simulations}

Based on the current knowledge on the BBH population, we provide simulation-based prospects for the upcoming LVK observing runs, simulating mock datasets at O4 and O5 sensitivities for the two mass models, \plg and \camel.
According to the official LVK plans~\cite{KAGRA:2013rdx}, we set $\sim 2$ yrs of observation time for O4 and $3$ yrs for O5, assuming realistic duty cycles for the detectors. 
We thus consider a four detector network for both observing run, including the two LIGO~\cite{LIGOScientific:2014pky}, the Virgo~\cite{VIRGO:2014yos} and the KAGRA interferometers~\cite{KAGRA:2020tym} (the latter joining the network late in O4).
Additional information can be found in App.~\ref{app:duty_cycles}.
We generate a synthetic population from the BBH rates corresponding to the maximum likelihood values obtained from O3 analyses with a fixed cosmology~\cite{Gennari:2025nho}.
The injected population parameters are reported in App.~\ref{app:priors}, alongside with the priors used in the hierarchical inference.
For our simulated data, we adopt the Plank 2015 results~\cite{Planck:2015fie} for a flat $\Lambda \rm CDM$ model.
Additional details can be found in App.~\ref{app:priors}.

We evaluate the detectability of each GW event based on its network matched-filter SNR (computed using \texttt{bilby}~\cite{Ashton:2018jfp, Romero-Shaw:2020owr, Smith:2019ucc}), setting the detection threshold at $\rm SNR = 12$.
Noise is simulated for each GW event in all detectors of the LVK network from the PSDs at O4 and O5 design sensisivities~\cite{KAGRA:2013rdx}, and the GW signal is modelled with the \texttt{IMRPhenomXHM} waveform~\cite{Garcia-Quiros:2020qpx}.
While the masses and the redshift follow the population distribution, the three-dimensional spin components are drawn uniformly in magnitude and isotropic in orientation.
This allows us to remain agnostic about the spins distribution, which is not included in our population models.
The remaining extrinsic parameters are also drawn uniformly from the same priors used for single event parameter estimation (PE), except for the geocentric time which is drawn uniformly over the entire observing run.
The priors on the single event parameters are further discussed in App.~\ref{app:single_events_PE}.
To make our analysis as realistic as possible, we perform full Bayesian PE on the detected events using \texttt{bilby}, to obtain the posterior samples required for the hierarchical inference.

Using \icarogw, we carry out joint hierarchical inference of the population and cosmological parameters with the simulated datasets generated with the procedure described above. 
We account for selection effects in the analysis by generating \textit{injections} which cover a sufficiently large portion of the parameter space, and assess their detectability with the same procedure used for the mock catalogues.
Details about the injections can be found in App.~\ref{app:injections}.
We adopt for the hierarchical inference the same population model used to generate the dataset.
By doing so, we assume any form of mismodelling in the population to be negligible.

\begin{table}[]
    \centering
    \begin{tabular*}{0.9\linewidth}{@{\extracolsep\fill}lcc }
        \toprule
        \multicolumn{3}{@{}c@{}}{\textbf{O4 simulated}} \\
        \textbf{Model} & $H_0 ~[\rm km \,s^{-1} \,Mpc^{-1}]$ & $\Omega_{\rm m, 0}$ \\
        \cmidrule{2-3}
        \plg & ${80}_{-24\,(36)}^{+39\,(73)}$ & ${0.38}_{-0.28\,(0.35)}^{+0.38\,(0.55)}$ \\
        \camel & ${62}_{-14\,(20)}^{+14\,(23)}$ & ${0.30}_{-0.22\,(0.28)}^{+0.39\,(0.58)}$ \\
        \midrule
        \multicolumn{3}{@{}c@{}}{\textbf{O5 simulated}} \\
        \textbf{Model} & $H_0 ~[\rm km \,s^{-1} \,Mpc^{-1}]$ & $\Omega_{\rm m, 0}$ \\
        \cmidrule{2-3}
        \plg & ${56}_{-8\,(11)}^{+12\,(20)}$ & ${0.48}_{-0.21\,(0.31)}^{+0.29\,(0.43)}$ \\
        \camel & ${68.2}_{-2.0\,(2.9)}^{+1.7\,(3.1)}$ & ${0.32}_{-0.04\,(0.06)}^{+0.06\,(0.08)}$ \\
        \bottomrule
    \end{tabular*}
    \caption{\justifying \footnotesize Median values of the inferred cosmological parameters for each analysis, with the associated 68\% (90\%) confidence intervals computed from their marginalized posterior distribution.}
    \label{tab:cosmo_pars_inferred_values}
\end{table}

The results are summarized in Fig.~\ref{fig:H0_posteriors}, where we show the marginalised posterior distributions on $H_0$.
The median values with their 68\% (90\%) C.I. are reported in Tab.~\ref{tab:cosmo_pars_inferred_values}.
As expected, the marginalized posteriors of the cosmological parameters are all compatible with the injected values at 90\% C.I.
From our O4 simulation, we find 146 (142) observed events up to redshift $z \sim 2$ which constrain $H_0$ with 20\% (40\%) accuracy, assuming a \camel (\plg) distribution. 
The results of our fiducial \plg analysis are in good agreement with other similar studies~\cite{Borghi:2023opd, Kiendrebeogo:2023hzf}.
However, we note that the \camel distribution significantly improves the constraints compared to the \plg by a factor $\roughly 2$. 
This improvement is starker than the one we obtained with real O3 data, where the constraints on $H_0$ with the \camel are improved by only 30\% with respect to the LVK \plg analysis.

\begin{figure}
    \centering
    \includegraphics[width=\linewidth]{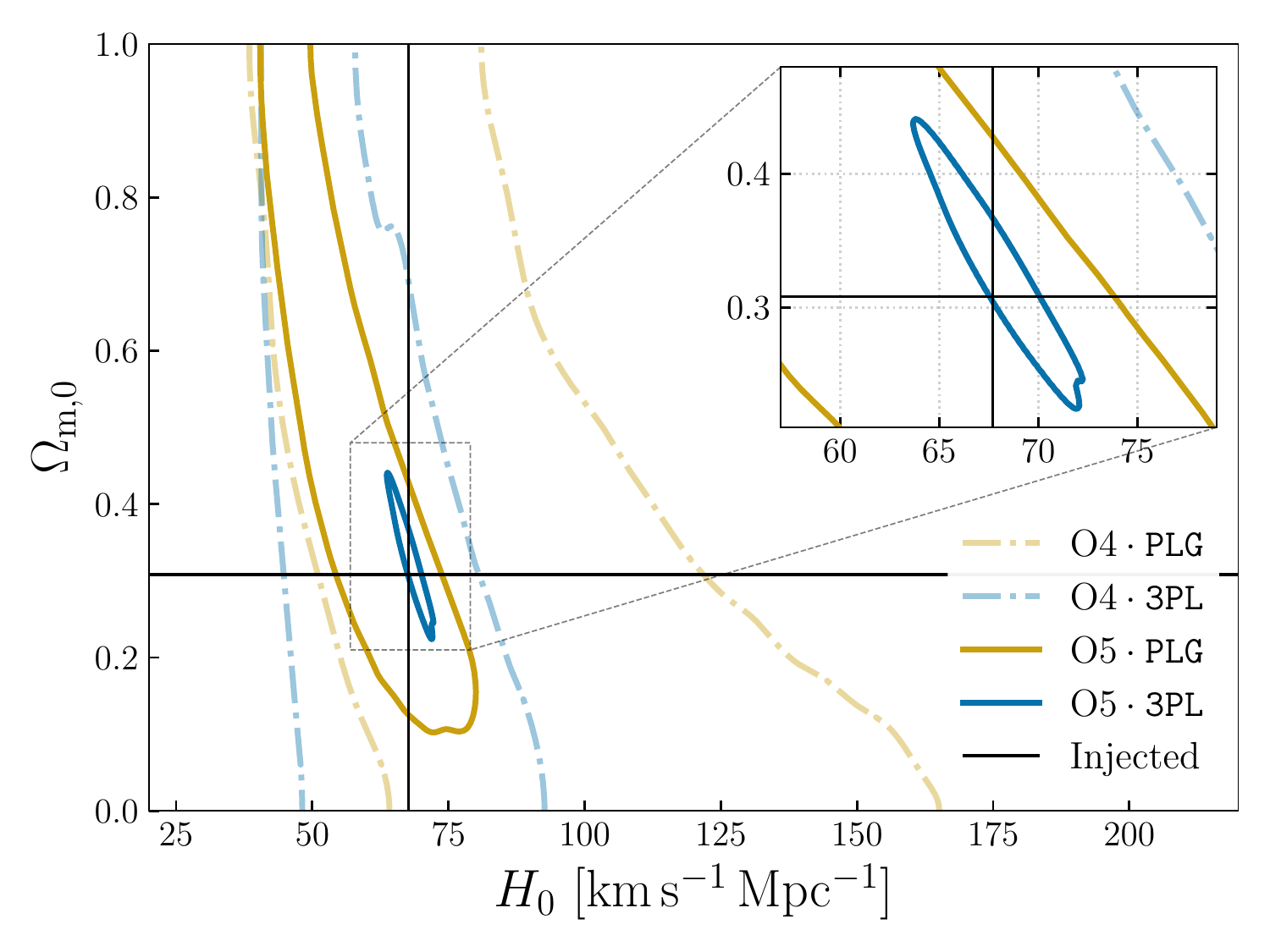}
    \caption{\justifying \footnotesize Marginalized posterior distributions on the cosmological parameters from O4 (dashed) and O5 (solid) simulated data, obtained with the \plg (yellow) and \camel (blue) population models. Contours correspond to $90\%$ credible regions. The injected values are indicated by the solid black line. The inset displays a magnified view of the parameter space, highlighting the O5 \camel contour. (\textit{Colors available online})}
    \label{fig:2d_cosmo}
\end{figure}

We observe a similar behaviour for O5, although much more pronounced. 
We find 1825 (1823) detected GW events up to redshift $z \sim 3$, constraining $H_0$ at 2.7\% (18\%) accuracy, assuming the \camel (\plg) distribution.
Our results suggest that the \camel model would perform more than 5 times better than the \plg by the end of O5 when it comes to measuring cosmological parameters with spectral sirens.
For a network of second-generation detectors, this result constitutes the most optimistic forecast of the Hubble constant constraints to date, using GW information only.
The additional striking finding of O5 \camel results is that the corresponding parametrization of the BBH population allows us to break the degeneracy between $H_0$ and $\Omega_{\rm m, 0}$, as clearly shown by Fig.~\ref{fig:2d_cosmo}.
The \camel thus not only provides accurate measurement of the Hubble constant, but also forecasts constraints with 15\% accuracy on the matter density parameter.

\section{Conclusion}
\label{sec:summary}

In this paper, we present a new GW-only measurement of the Hubble constant using a novel BBH primary mass distribution with three distinctive features (\camel), that is preferred to the fiducial \plg by LVK O3 data.
We improve GWTC-3 constraints on $H_0$ from GWs alone by a factor $\roughly 1.5$ compared to previous LVK analyses~\cite{LIGOScientific:2021aug}, with a measurement of $H_0 = 84^{+35}_{-25} ~\rm km \,s^{-1} \,Mpc^{-1}$.

Our results are comparable with the constraints recently released by the LVK collaboration using their latest GWTC-4 catalog~\cite{LIGOScientific:2025hdt, LIGOScientific:2025slb, LIGOScientific:2025yae}, including events from the first part of the ongoing fourth observing run (O4a).
In the context of spectral sirens, the latest constraints on the Hubble constant from GWTC-4 are $H_0 = 77.1^{+40.8}_{-26.3} ~\rm km \,s^{-1} \,Mpc^{-1}$ inferred from BBH events only with a mass model with a power-law and two Gaussian peaks, and $H_0 = 76.4^{+23.0}_{-18.1} ~\rm km \,s^{-1} \,Mpc^{-1}$ when including neutron star events~\cite{LIGOScientific:2025pvj, LIGOScientific:2025jau}.
These correspond to accuracy of 44\% and 27\% respectively.
Our \camel measurement, achieving 35\% accuracy with only 50 BBH events from O3, are well comparable to these results, although relying on almost three times less events.
A thorough analysis of GWTC-4.0 data with the \camel model requires a full comparison with the population models presented in \cite{LIGOScientific:2025jau}, as well as the addition of neutron star events, and will be presented in a future publication.
%
%

%
In general, our results highlights how properly resolving intrinsic structure in the BBH population can dramatically enhance cosmological inference with GWs.
By projecting the \camel model to upcoming LVK observing runs, we provide forecasts for the constraints on $H_0$ and $\Omega_{m,0}$ in the near future.
We report an expected constraint on $H_0$ at 2.7\% by the end of O5.
Such a result constitutes the tightest forecast on $H_0$ so far estimated for O5, showing that a GW solution of the Hubble tension is within reach of the current second-generation detector network, even without relying on complementary EM information.

The results obtained in our investigation strongly depend on the \camel BBH population model we considered, which presents three sharp features in the primary mass distribution driving the precise cosmological inference.
Even if \camel is preferred by O3 data, nothing guarantees that it will be still favoured by future observations (this point in particular is of crucial interest when it comes to future GWTC-4.0 analysis).
Nevertheless if sharp features in the BBH mass distribution similar to the ones of \camel are indeed present, even if another model will be preferred we may expect similar cosmological constraints \cite{Farr:2019twy}.

Furthermore \camel assumes a stationary, i.e.~not evolving in redshift, BBH mass distribution.
An unmodelled redshift evolution of the BBH mass function may provide a substantial cosmological bias, if not properly taken into account~\cite{Mukherjee:2021rtw, Pierra:2023deu, Agarwal:2024hld, Mali:2024wpq}.
However the observed mass distribution does not show any evidence of redshift evolution so far \cite{Karathanasis:2022rtr, Lalleman:2025xcs, Gennari:2025nho}, and an improved \camel model capable of correctly capturing potential redshift evolution of one or more features, is not expected to considerably degrade the results provided here, barring extreme deviations from stationarity at high redshift.

Finally, future GW observations could actually not provide the number of detections considered in our analyses (e.g.~because of a lower BBH merger rate or a shorter observational time), or the preferred population model results to be one with (slightly) broader mass features.
In such cases, we can expect a moderate degradation of the cosmological results presented here.
Such a loss could however be compensated by the use of complementary EM information, either in the form of galaxy catalogs or by the observations of more EM counterparts.

Even considering all these possible issues, the general conclusion we can draw from our investigation is that a solution of the Hubble tension is within reach of the current network of GW detectors.

%
%
%
%
%
%

\section*{Acknowledgements}

We thank Sylvain Marsat for his valuable advice at early stages of the project and Alexander Papadopoulos for insights on the simulations.
We thank Simone Mastrogiovanni and Danièle Steer for useful discussions on the simulation pipeline setups.
We are grateful to the LIGO-Virgo-KAGRA Cosmology group for stimulating discussion and to Gregoire Pierra for serving as internal review for the LIGO-Virgo-KAGRA Collaboration (document number: P2500486).
T.B., V.G, N.T.~acknowledge support from the French space agency CNES in the framework of LISA.
T.B. acknowledges support from a CDSN PhD grant from ENS-PSL.
This project has received financial support from the CNRS through the AMORCE funding framework and from the Agence Nationale de la Recherche (ANR) through the MRSEI project ANR-24-MRS1-0009-01.
The authors are grateful for computational resources provided by the IN2P3 computing centre (CC-IN2P3) in Lyon (Villeurbanne).
This research made use of data, software and/or web tools obtained from the Gravitational Wave Open Science Center~\cite{LIGOScientific:2019lzm, KAGRA:2023pio}, a service of the LIGO Scientific Collaboration, the KAGRA Collaboration and the Virgo Collaboration.
This material is based upon work supported by NSF's LIGO Laboratory which is a major facility fully funded by the National Science Foundation.
LIGO Laboratory and Advanced LIGO are funded by the United States National Science Foundation (NSF) as well as the Science and Technology Facilities Council (STFC) of the United Kingdom, the Max-Planck Society (MPS), and the State of Niedersachsen/Germany for support of the construction of Advanced LIGO and construction and operation of the GEO600 detector. Additional support for Advanced LIGO was provided by the Australian Research Council. Virgo is funded, through the European Gravitational Observatory (EGO), by the French Centre National de Recherche Scientifique (CNRS), the Italian Istituto Nazionale di Fisica Nucleare (INFN) and the Dutch Nikhef, with contributions by institutions from Belgium, Germany, Greece, Hungary, Ireland, Japan, Monaco, Poland, Portugal, Spain. KAGRA is supported by Ministry of Education, Culture, Sports, Science and Technology (MEXT), Japan Society for the Promotion of Science (JSPS) in Japan; National Research Foundation (NRF) and Ministry of Science and ICT (MSIT) in Korea; Academia Sinica (AS) and National Science and Technology Council (NSTC) in Taiwan.\\

\noindent \textbf{Competing interest} The authors declare no competing interests.

\noindent \textbf{Authors contribution} T.B. and V.G. developped the simulation pipeline, processed the data, and produced the results. N.T supervised the project.
All authors contributed to the writing of the paper.

\noindent \textbf{Correspondence and requests for materials} Queries should be addressed to T.B. or V.G.

\section*{Appendix}
\appendix
\section{Detectors duty cycles}
\label{app:duty_cycles}

We report in table~\ref{tab:duty_cycles_bns_range} the duty cycles as well as the binary neutron star (BNS) ranges of the detectors (i.e. their sensitivities) assumed for O4 and O5 simulations.
All relevant data is taken from Ref.~\cite{KAGRA:2013rdx} and is publicly available.
Note that for O4, Virgo and KAGRA duty cycles are well below LIGO's. This reflects the fact that they have joined the observation run later than the starting date.
For O5, we assume realistic uniform duty cycles of $70\%$ across all detectors.
We assume O4 spans from May 24th 2023 to 0ctober 7th 2025,
with two $\roughly 1$ month long commissioning breaks, hence a total observation time of a bit less than 2 years.
We assume O5 spans from January 1st 2028 to December 31st 2030, hence a total observation time of 3 years.

\begin{table}[ht!]
    \centering
    \begin{tabular*}{\linewidth}{@{\extracolsep\fill}lcccccccc}
        \toprule%
        & \multicolumn{4}{@{}c@{}}{\textbf{Duty cycle}} & \multicolumn{4}{@{}c@{}}{\textbf{BNS range} [Mpc]} \\\cmidrule{2-5}\cmidrule{6-9}%
         & H & L & V & K & H & L & V & K \\
        \midrule
        \textbf{O4} & $65\%$ & $80\%$ & $47\%$ & $15\%$ & $190$ & $190$ & $120$ & $10$ \\
        \textbf{O5} & $70\%$ & $70\%$ & $70\%$ & $70\%$ & $330$ & $330$ & $150$ & $80$ \\
        \bottomrule
    \end{tabular*}
    \caption{\justifying \footnotesize Duty cycles (left) and approximate BNS ranges (right) used in the simulations for the four interferometers of the current network of GW detectors: LIGO Handford (H), LIGO Livingstone (L), Virgo (V), KAGRA (K).}
    \label{tab:duty_cycles_bns_range}
\end{table}

\section{Population models and priors}
\label{app:priors}

\begin{figure}
    \centering
    \includegraphics[width=\linewidth]{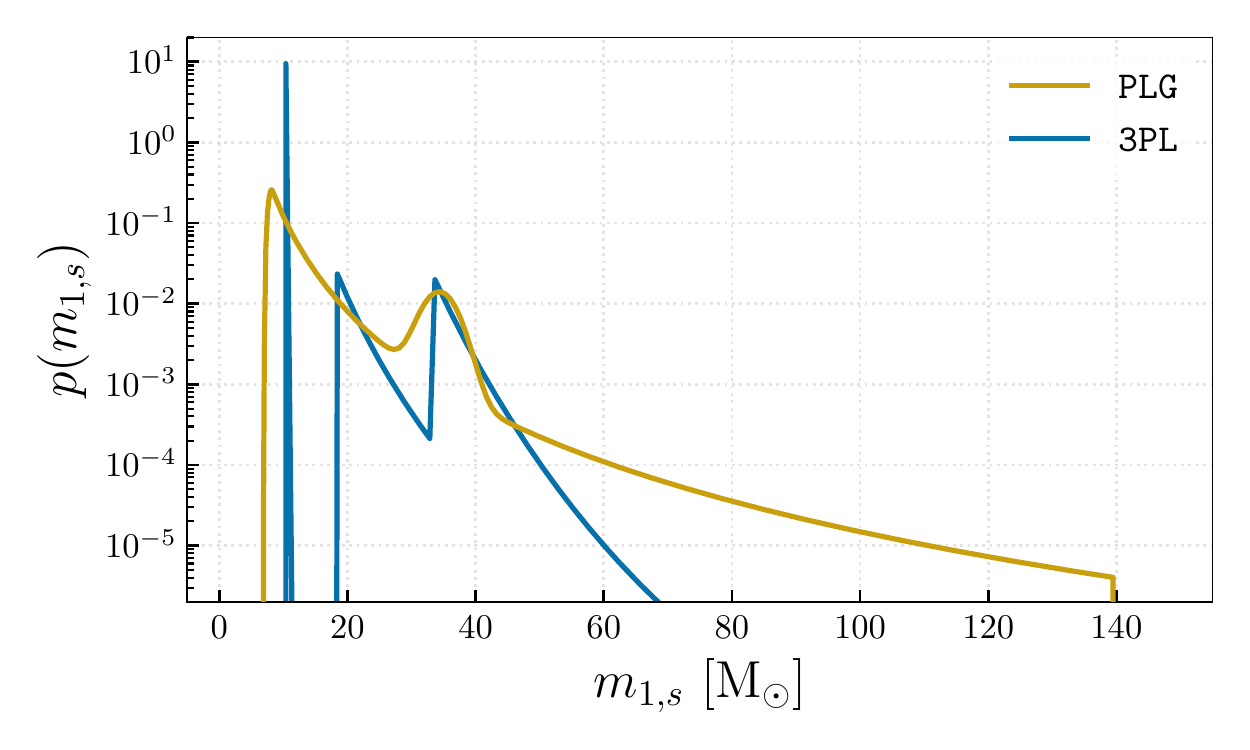}
    \caption{\justifying \footnotesize Source frame primary mass distributions injected in the simulations. The population parameters correspond to the maximum likelihood values found in O3 analyses with fixed cosmology~\cite{Gennari:2025nho}. The fact that the \camel model describes a more intricate population with additional features compared to the \plg model, as well as the sharpness with which it captures the first feature at $\sim 10~\rm{M_\odot}$ is clear in this plot. (\textit{Colors available online})}
    \label{fig:primary_mass_injected}
\end{figure}

The reader can refer to Ref.~\cite{Gennari:2025nho} for a complete description of the population models used in the present work.
We list in Tab.~\ref{tab:models_priors} the priors used for all the hyperparameters.
The priors for O3 analyses--reported in the first column--are taken from Ref.~\cite{Gennari:2025nho} for the \camel, while LVK-like priors~\cite{LIGOScientific:2021aug} are used for the \plg (although wider for $m_{\rm min}$ and $\mu$).
The second column reports the maximum likelihood values from O3 analyses with fixed cosmology~\cite{Gennari:2025nho}, which we use as the injected values in our simulations.
The shape of the injected primary mass distributions is shown in Fig.~\ref{fig:primary_mass_injected}.
The priors used in the O4 and O5 simulations--reported in the last column--slightly differ from the ones used on real data, taking advantage of the narrower posteriors induced by the larger number of events compared to O3.
We make sure that the posteriors support is largely contained within the prior bounds.

Differently than on real data, we use a \textit{scale-free} likelihood for the simulations, in which the local event rate $R_0$ is analytically marginalised over assuming a Jeffrey's prior (e.g. see~\cite{Mastrogiovanni:2023zbw}).
This allows us to reduce the dimensionality of the parameter space and consequently the computation time.

For O3 analyses, we use the \dynesty sampler~\cite{Speagle:2019ivv, dynesty} from \bilby~\cite{Ashton:2018jfp, Romero-Shaw:2020owr} which implements a nested sampling algorithm~\cite{Skilling2006NestedSF} to sample the hierarchical likelihood.
To reduce the computational time, we use the \nessai sampler~\cite{nessai, Williams:2021qyt, Williams:2023ppp} for O4 and O5 simulations, that implements a fast nested sampling which exploits normalizing flows to approximate the likelihood that is used to move live points.
For both samplers, we use 2000 live points to accurately explore the complex hierarchical likelihood, and use $\texttt{dlogZ} = 0.1$ as stopping criterion.

\begin{table*}
    \begin{tabular*}{0.9\textwidth}{@{\extracolsep\fill}lccc}
    
        \toprule%
        
        \multirow{2}{*}{\textbf{Models \& parameters}} & \multicolumn{1}{@{}c@{}}{\textbf{O3 real data}} & \multicolumn{2}{@{}c@{}}{\textbf{O4 \& O5 simulations}} \\\cmidrule{2-2}\cmidrule{3-4}%
        
         & priors & injected (O3 $\max \mathcal{L})$ & priors \\
        \midrule
        
        \textbf{Primary mass} & & & \\\cmidrule{1-1}
        
        \plg &&& \\
        \hspace{5pt} $\alpha$ & (-4, 12) & 3.91 & (2, 6) \\
        \hspace{5pt} $m_{\rm min} ~\rm[M_\odot]$ & (1, 100) & 6.73 & (1, 10) \\
        \hspace{5pt} $m_{\rm max} ~\rm[M_\odot]$ & (30, 200) & 140 & (80, 200) \\
        \hspace{5pt} $\delta_m ~\rm[M_\odot]$ & (0, 10) & 1.82 & - \\
        \hspace{5pt} $\mu ~\rm[M_\odot]$ & (1, 60) & 34.3 & (20, 50) \\
        \hspace{5pt} $\sigma ~\rm[M_\odot]$ & (1, 6) & 2.59 & (0.4, 10) \\
        \hspace{5pt} ${\rm mix}$ & (0, 1) & 0.91 & - \\[5pt]
        
        \camel &&& \\
        \hspace{5pt} $\alpha_a$ & (-4, 200) & 181.2 & (20, 250) \\
        \hspace{5pt} $m_{{\rm min}, a} ~\rm[M_\odot]$ & (1, 15) & 10.36 & (8, 13) \\
        \hspace{5pt} $m_{{\rm max}, a} ~\rm[M_\odot]$ & (30, 200) & 44.5 & (30, 60) \\
        \hspace{5pt} $\alpha_b$ & (-4, 50) & 8.1 & (0, 30) \\
        \hspace{5pt} $m_{{\rm min}, b} ~\rm[M_\odot]$ & (10, 20) & 17.8 & (13, 25) \\
        \hspace{5pt} $m_{{\rm max}, b} ~\rm[M_\odot]$ & (30, 200) & 124 & (110, 140) \\
        \hspace{5pt} $\alpha_c$ & (-4, 50) & 13.4 & (0, 30) \\
        \hspace{5pt} $m_{{\rm min}, c} ~\rm[M_\odot]$ & (15, 60) & 33.4 & (25, 50) \\
        \hspace{5pt} $m_{{\rm max}, c} ~\rm[M_\odot]$ & (30, 200) & 172 & (155, 185) \\
        \hspace{5pt} ${\rm mix}_a$ & (0, 1) & 0.86 & - \\
        \hspace{5pt} ${\rm mix}_b$ & (0, 1) & 0.08 & - \\
        \midrule
        
        \textbf{Mass ratio} & & & \\\cmidrule{1-1}
        
        \texttt{Power-law} &&& \\
        \hspace{5pt} $\alpha_q$ & (-20, 20) & 4.7 & (0, 8) \\[5pt]
        
        \texttt{Gaussian} &&& \\
        \hspace{5pt} $\mu_q$ & (0.1, 1) & 0.78 & - \\
        \hspace{5pt} $\sigma_q$ & (0.05, 0.9) & 0.17 & - \\
        \midrule
        
        \textbf{Rate evolution} & & & \\\cmidrule{1-1}

        \hspace{5pt} $R_0~\rm[yr^{-1}\,Gpc^{-3}] $ & (0, 100) & 19.8 (\plg) / 23.7 (\camel) & Not inferred \\[5pt]
        
        \texttt{MadauDickinson} &&& \\
        \hspace{5pt} $\gamma$ & (-50, 30) & 1.8 (\plg) / 4.5 (\camel) & (-10, 15) \\
        \hspace{5pt} $\kappa$ & (-20, 10) & 5.9 (\plg) / -0.05 (\camel) & (-10, 10) \\
        \hspace{5pt} $z_p$ & (0, 4) & 1.9 (\plg) / 0.33 (\camel) & - \\
        \midrule
        
        \textbf{Cosmology} & & & \\\cmidrule{1-1}
        
        \texttt{FlatLambdaCDM} &&& \\
        \hspace{5pt} $H_0 ~[\rm km \,s^{-1} \,Mpc^{-1}]$ & (10, 200) & 67.7 & - \\
        \hspace{5pt} $\Omega_{\rm m, 0}$ & (0, 1) & 0.308 & - \\
        
        \bottomrule
    \end{tabular*}
    \caption{\justifying \footnotesize Priors for the population parameters of the different rate models considered. For every parameter, a uniform prior in the indicated range is employed. Empty entries in the last column \textit{simulation priors} means that the same prior ranges as in O3 hierarchical inference have been used. Note that the injected values shown here are rounded, except for the cosmological parameters. A visual representation of the injected primary mass distributions for the simulations is shown in Fig.\ref{fig:primary_mass_injected}.}
    \label{tab:models_priors}
\end{table*}

\section{Events parameter estimation}
\label{app:single_events_PE}

For each of the \textit{detected} event in our simulated datasets, we estimate their parameters using \bilby~\cite{Ashton:2018jfp, Romero-Shaw:2020owr, Smith:2019ucc}.
We simulate colored Gaussian noise in the detectors and consequently employ a Gaussian likelihood in frequency domain, bandpassed between $[20, 1024]~{\rm Hz}$.
We use the \texttt{IMRPhenomXHM} waveform model~\cite{Garcia-Quiros:2020qpx} which includes higher angular harmonics but neglects precession (thus assuming aligned spins binaries).
We generate the signal and noise at sampling frequency $2048~{\rm Hz}$, starting from $20~{\rm Hz}$.
The signal duration is $t_{\rm seglen} = \Delta t_{\rm mrg} + 4~{\rm s}$, with $\Delta t_{\rm mrg}$ being the time difference between the moment the binary's frequency enters the detector's band ($20~{\rm Hz}$) and the frequency at merger.
The analysis start time is set to be at $t_{\rm seglen}+1~{\rm s}$ before the geocent time of the event (see below).

We sample over the following 11 parameters: $m_1, m_2$ (primary and secondary masses, such that $m_1 > m_2$), $d_L$ (luminosity distance), $\theta_{JN}$ (inclination angle), $\psi$ (polarization angle), $\phi$ (phase of the signal), $\alpha, \delta$ (right ascension and declination, i.e.~the sky localization of the binary), $\chi_1, \chi_2$ (projected component of the spin vectors of each BH along the orbital angular momentum) and $t_{\rm geocent}$ (the time at which the merger signal reaches the Earth's center).

We employ uniform priors for the component masses $m_1$ and $m_2$, within a range mildly tailored to the injected value $m_{1, \rm inj}$, i.e. between $1~\rm M_\odot$ and an upper bound linearly interpolated from the following reference values: $60~\rm M_\odot (300~\rm M_\odot)$ for $m_{1, \rm inj} = 10~\rm M_\odot (90~\rm M_\odot)$.
We employ a uniform prior for the luminosity distance $d_L$ in the range $[0, 16]~{\rm Gpc}$ for O4-like datasets and $[0, 30]~{\rm Gpc}$ for O5-like datasets.
We assume a spin prior uniform in magnitude and orientation (see Eq.~A7 of \cite{Lange:2018pyp}), similar to what is done for real events \cite{LIGOScientific:2018mvr, LIGOScientific:2021usb, KAGRA:2021vkt}.
For the sky localization we use a prior uniform on the celestial sphere.
The prior on $\cos \theta_{JN}$ is uniform in $[-1, 1]$ so that the prior on the binary orientation is uniform on the unit sphere.
The priors for $\phi, \psi$ is uniform in $[0, 2\pi]$.
Finally, the prior for $t_{\rm geocent}$ is uniform in the range $[t_{\rm g, inj} - 0.1~{\rm s}, t_{\rm g, inj} + 0.1~{\rm s}]$ where $t_{\rm g, inj}$ is the injected geocentric time value. 

We sample the likelihood with the \nessai sampler~\cite{nessai, Williams:2021qyt, Williams:2023ppp}.
To ensure a stable PE across the entire parameter space, we use 5000 live points for each event.
If 2000 live points would be sufficient for most events, we found the sampler to struggle with low mass events, a problem which was solved by increasing the number of live points.
Note that the computational cost was only mildly impacted by this choice.
As for the hierarchical inference, we use $\texttt{dlogZ} < 0.1$ as the stopping criterion.

\section{Evaluating selection effects}
\label{app:injections}

Population analyses with GWs are heavily influenced by \textit{selection effects} due to the finite detectors sensitivity~\cite{Gaebel:2018poe, Mandel:2020cig, Kapadia:2019uut, Vitale:2020aaz}. 
This effect is directly taken into account in the likelihood, by generating a large set of simulated events--normally called \textit{injections}--and evaluate their detectability in realistic detector noise~\cite{Essick:2023upv}.
For O3 analyses, we use the publicly available injection by LVK computed at O3 sensitivity in real noise~\cite{O3:bbhpop}, that have been used in the LVK publications for GWTC-3~\cite{KAGRA:2021duu, LIGOScientific:2021aug}. 
For simulations, we generate our own injection sets at O4 and O5 sensitivities, following a procedure similar to Ref.~\cite{Agarwal:2024hld}.
We draw events from a wide population distribution that largely covers the target one, to ensure that a sufficient number of points is available to evaluate selection effects by Monte Carlo integration during the inference (e.g. see~\cite{Mastrogiovanni:2023zbw}).
We adopt a truncated power-law with index $\alpha = - 3$ for the primary mass between $2~{\rm M_\odot}$ and $150~{\rm M_\odot}$, smoothed at its lower bound with a wide tempering of $\delta_m = 17~{\rm M_\odot}$; a truncated power-law with index $\alpha_q = 1$ for the mass ratio; a power-law with index $\gamma = 0$ for the rate evolution.
To consistently estimate the detector sensitivity, we process the simulated injections through the same detection procedure as the events in our simulated datasets, i.e. computing the matched-filter SNR in colored Gaussian noise and labelling as detected those above the $\rm SNR > 12$ threshold, as described in Sec.~\ref{sec:simulations}.
Following the aforementioned method, we generate $5 \cdot 10^5$ ($6 \cdot 10^6$) detected injections for O4 (O5) simulations.
We comment on the numerical stability in App.~\ref{app:numerical_stability}.

\section{Numerical stability}\label{app:numerical_stability}

The hierarchical likelihood is numerically evaluated via Monte Carlo integrals~\cite{Mastrogiovanni:2023zbw}.
To ensure numerical stability of such integrals, we employ the \textit{effective number of PE samples} $\texttt{Neff}_{\texttt{PE}}$ and the \textit{effective number of injections} $\texttt{Neff}_{\texttt{inj}}$ as stability estimators~\cite{Mastrogiovanni:2023zbw}, and restrict the hierarchical likelihood parameter space in the regions where $\texttt{Neff}_{\texttt{PE}} >10$ and $\texttt{Neff}_{\texttt{inj}} > 4 \cdot \texttt{N}_{\texttt{obs}}$, with $\texttt{N}_{\texttt{obs}}$ the number of observed events.
We use 2048 PE samples for each O3 event, while $10^4$ for O4 and O5 forecasts.
The number of injections used for selection effects is reported in App.~\ref{app:injections}.
In Tab.~\ref{tab:numerical_stability}, we report the numerical stability estimators for the inferred population maximum likelihood values.
For O4 and O5 simulations, we find those values to be in agreement with $\texttt{Neff}_{\texttt{PE}}$ and $\texttt{Neff}_{\texttt{inj}}$ corresponding to the injected population.

\begin{table}[ht!]
    \centering
    \begin{tabular*}{0.9\linewidth}{@{\extracolsep\fill}lcccccc}
        \toprule
        \multirow{2}{*}{\textbf{Model}} & \multicolumn{3}{@{}c@{}}{$\texttt{Neff}_{\texttt{PE}}$} & \multicolumn{3}{@{}c@{}}{$\texttt{Neff}_{\texttt{inj}} / 4 \cdot \texttt{N}_{\texttt{obs}}$}\\\cmidrule{2-4}\cmidrule{5-7}
        & O3 & O4 & O5 & O3 & O4 & O5 \\
        \midrule
        \camel & 13 & 47 & 11  & 9  & 6   & 7  \\
        \plg   & 27 & 94 & 140 & 31 & 218 & 237 \\
        \bottomrule
    \end{tabular*}
    \caption{\justifying \footnotesize Effective number of PE samples and injections for the different analyses, evaluated at the maximum likelihood population samples.}
    \label{tab:numerical_stability}
\end{table}

In our analysis, the requirements above need to be carefully addressed, given the sharp features involved in the \camel population.
In Ref.~\cite{Gennari:2025nho}, the stability of the results was tested under different stability criteria, especially using a threshold on the log-likelihood variance to be less than $1$, as recommended in Ref.~\cite{Talbot:2023pex}.
For this reason, we use the same criteria on the effective samples and injections adopted in Ref.~\cite{Gennari:2025nho}, for both O3 data and simulations.
We note, however, for O4 and O5 simulations involving a large number of events, that the $10 M_{\odot}$ peak of the \camel is prone to numerical instability from $\texttt{Neff}_{\texttt{PE}}$, driven by the few events whose noise realization statistically shifts $m_1$ posteriors to lower values, leaving $10 M_{\odot}$ in the tail of the distribution.
Such few events have been removed from the analysis, but found not to affect the results.
Specifically, for the \camel simulated populations, we removed 3 (37) such events from the O4 (O5) hierarchical analyses, which in both scenario represent $\roughly 2\%$ of the total number of detected events in the dataset.
By generating several realizations of the same population, we checked that such variation is negligible compared to the Poisson noise from the events draws.
These limitations concerning the \camel model will hopefully disappear once a better parametrization for the primary mass is found.

\bibliography{references}

\begin{thebibliography}{87}%
\makeatletter
\providecommand \@ifxundefined [1]{%
 \@ifx{#1\undefined}
}%
\providecommand \@ifnum [1]{%
 \ifnum #1\expandafter \@firstoftwo
 \else \expandafter \@secondoftwo
 \fi
}%
\providecommand \@ifx [1]{%
 \ifx #1\expandafter \@firstoftwo
 \else \expandafter \@secondoftwo
 \fi
}%
\providecommand \natexlab [1]{#1}%
\providecommand \enquote  [1]{``#1''}%
\providecommand \bibnamefont  [1]{#1}%
\providecommand \bibfnamefont [1]{#1}%
\providecommand \citenamefont [1]{#1}%
\providecommand \href@noop [0]{\@secondoftwo}%
\providecommand \href [0]{\begingroup \@sanitize@url \@href}%
\providecommand \@href[1]{\@@startlink{#1}\@@href}%
\providecommand \@@href[1]{\endgroup#1\@@endlink}%
\providecommand \@sanitize@url [0]{\catcode `\\12\catcode `\$12\catcode `\&12\catcode `\#12\catcode `\^12\catcode `\_12\catcode `\%12\relax}%
\providecommand \@@startlink[1]{}%
\providecommand \@@endlink[0]{}%
\providecommand \url  [0]{\begingroup\@sanitize@url \@url }%
\providecommand \@url [1]{\endgroup\@href {#1}{\urlprefix }}%
\providecommand \urlprefix  [0]{URL }%
\providecommand \Eprint [0]{\href }%
\providecommand \doibase [0]{http://dx.doi.org/}%
\providecommand \selectlanguage [0]{\@gobble}%
\providecommand \bibinfo  [0]{\@secondoftwo}%
\providecommand \bibfield  [0]{\@secondoftwo}%
\providecommand \translation [1]{[#1]}%
\providecommand \BibitemOpen [0]{}%
\providecommand \bibitemStop [0]{}%
\providecommand \bibitemNoStop [0]{.\EOS\space}%
\providecommand \EOS [0]{\spacefactor3000\relax}%
\providecommand \BibitemShut  [1]{\csname bibitem#1\endcsname}%
\let\auto@bib@innerbib\@empty
\bibitem [{\citenamefont {Abbott}\ \emph {et~al.}(2016{\natexlab{a}})\citenamefont {Abbott} \emph {et~al.}}]{LIGOScientific:2016aoc}%
  \BibitemOpen
  \bibfield  {author} {\bibinfo {author} {\bibfnamefont {B.~P.}\ \bibnamefont {Abbott}} \emph {et~al.} (\bibinfo {collaboration} {LIGO Scientific, Virgo}),\ }\bibfield  {title} {\enquote {\bibinfo {title} {{Observation of Gravitational Waves from a Binary Black Hole Merger}},}\ }\href {\doibase 10.1103/PhysRevLett.116.061102} {\bibfield  {journal} {\bibinfo  {journal} {Phys. Rev. Lett.}\ }\textbf {\bibinfo {volume} {116}},\ \bibinfo {pages} {061102} (\bibinfo {year} {2016}{\natexlab{a}})},\ \Eprint {http://arxiv.org/abs/1602.03837} {arXiv:1602.03837 [gr-qc]} \BibitemShut {NoStop}%
\bibitem [{\citenamefont {Abbott}\ \emph {et~al.}(2019{\natexlab{a}})\citenamefont {Abbott} \emph {et~al.}}]{LIGOScientific:2018mvr}%
  \BibitemOpen
  \bibfield  {author} {\bibinfo {author} {\bibfnamefont {B.~P.}\ \bibnamefont {Abbott}} \emph {et~al.} (\bibinfo {collaboration} {LIGO Scientific, Virgo}),\ }\bibfield  {title} {\enquote {\bibinfo {title} {{GWTC-1: A Gravitational-Wave Transient Catalog of Compact Binary Mergers Observed by LIGO and Virgo during the First and Second Observing Runs}},}\ }\href {\doibase 10.1103/PhysRevX.9.031040} {\bibfield  {journal} {\bibinfo  {journal} {Phys. Rev. X}\ }\textbf {\bibinfo {volume} {9}},\ \bibinfo {pages} {031040} (\bibinfo {year} {2019}{\natexlab{a}})},\ \Eprint {http://arxiv.org/abs/1811.12907} {arXiv:1811.12907 [astro-ph.HE]} \BibitemShut {NoStop}%
\bibitem [{\citenamefont {Abbott}\ \emph {et~al.}(2024)\citenamefont {Abbott} \emph {et~al.}}]{LIGOScientific:2021usb}%
  \BibitemOpen
  \bibfield  {author} {\bibinfo {author} {\bibfnamefont {R.}~\bibnamefont {Abbott}} \emph {et~al.} (\bibinfo {collaboration} {LIGO Scientific, VIRGO}),\ }\bibfield  {title} {\enquote {\bibinfo {title} {{GWTC-2.1: Deep extended catalog of compact binary coalescences observed by LIGO and Virgo during the first half of the third observing run}},}\ }\href {\doibase 10.1103/PhysRevD.109.022001} {\bibfield  {journal} {\bibinfo  {journal} {Phys. Rev. D}\ }\textbf {\bibinfo {volume} {109}},\ \bibinfo {pages} {022001} (\bibinfo {year} {2024})},\ \Eprint {http://arxiv.org/abs/2108.01045} {arXiv:2108.01045 [gr-qc]} \BibitemShut {NoStop}%
\bibitem [{\citenamefont {Abbott}\ \emph {et~al.}(2021{\natexlab{a}})\citenamefont {Abbott} \emph {et~al.}}]{LIGOScientific:2020ibl}%
  \BibitemOpen
  \bibfield  {author} {\bibinfo {author} {\bibfnamefont {R.}~\bibnamefont {Abbott}} \emph {et~al.} (\bibinfo {collaboration} {LIGO Scientific, Virgo}),\ }\bibfield  {title} {\enquote {\bibinfo {title} {{GWTC-2: Compact Binary Coalescences Observed by LIGO and Virgo During the First Half of the Third Observing Run}},}\ }\href {\doibase 10.1103/PhysRevX.11.021053} {\bibfield  {journal} {\bibinfo  {journal} {Phys. Rev. X}\ }\textbf {\bibinfo {volume} {11}},\ \bibinfo {pages} {021053} (\bibinfo {year} {2021}{\natexlab{a}})},\ \Eprint {http://arxiv.org/abs/2010.14527} {arXiv:2010.14527 [gr-qc]} \BibitemShut {NoStop}%
\bibitem [{\citenamefont {Abbott}\ \emph {et~al.}(2023{\natexlab{a}})\citenamefont {Abbott} \emph {et~al.}}]{KAGRA:2021vkt}%
  \BibitemOpen
  \bibfield  {author} {\bibinfo {author} {\bibfnamefont {R.}~\bibnamefont {Abbott}} \emph {et~al.} (\bibinfo {collaboration} {KAGRA, VIRGO, LIGO Scientific}),\ }\bibfield  {title} {\enquote {\bibinfo {title} {{GWTC-3: Compact Binary Coalescences Observed by LIGO and Virgo during the Second Part of the Third Observing Run}},}\ }\href {\doibase 10.1103/PhysRevX.13.041039} {\bibfield  {journal} {\bibinfo  {journal} {Phys. Rev. X}\ }\textbf {\bibinfo {volume} {13}},\ \bibinfo {pages} {041039} (\bibinfo {year} {2023}{\natexlab{a}})},\ \Eprint {http://arxiv.org/abs/2111.03606} {arXiv:2111.03606 [gr-qc]} \BibitemShut {NoStop}%
\bibitem [{\citenamefont {Abbott}\ \emph {et~al.}(2021{\natexlab{b}})\citenamefont {Abbott} \emph {et~al.}}]{LIGOScientific:2020kqk}%
  \BibitemOpen
  \bibfield  {author} {\bibinfo {author} {\bibfnamefont {R.}~\bibnamefont {Abbott}} \emph {et~al.} (\bibinfo {collaboration} {LIGO Scientific, Virgo}),\ }\bibfield  {title} {\enquote {\bibinfo {title} {{Population Properties of Compact Objects from the Second LIGO-Virgo Gravitational-Wave Transient Catalog}},}\ }\href {\doibase 10.3847/2041-8213/abe949} {\bibfield  {journal} {\bibinfo  {journal} {Astrophys. J. Lett.}\ } (\bibinfo {year} {2021}{\natexlab{b}}),\ 10.3847/2041-8213/abe949},\ \Eprint {http://arxiv.org/abs/2010.14533} {arXiv:2010.14533} \BibitemShut {NoStop}%
\bibitem [{\citenamefont {Abbott}\ \emph {et~al.}(2023{\natexlab{b}})\citenamefont {Abbott} \emph {et~al.}}]{KAGRA:2021duu}%
  \BibitemOpen
  \bibfield  {author} {\bibinfo {author} {\bibfnamefont {R.}~\bibnamefont {Abbott}} \emph {et~al.} (\bibinfo {collaboration} {KAGRA, VIRGO, LIGO Scientific}),\ }\bibfield  {title} {\enquote {\bibinfo {title} {{Population of Merging Compact Binaries Inferred Using Gravitational Waves through GWTC-3}},}\ }\href {\doibase 10.1103/PhysRevX.13.011048} {\bibfield  {journal} {\bibinfo  {journal} {Phys. Rev. X}\ } (\bibinfo {year} {2023}{\natexlab{b}}),\ 10.1103/PhysRevX.13.011048},\ \Eprint {http://arxiv.org/abs/2111.03634} {arXiv:2111.03634} \BibitemShut {NoStop}%
\bibitem [{\citenamefont {Abbott}\ \emph {et~al.}(2017{\natexlab{a}})\citenamefont {Abbott} \emph {et~al.}}]{LIGOScientific:2017adf}%
  \BibitemOpen
  \bibfield  {author} {\bibinfo {author} {\bibfnamefont {B.~P.}\ \bibnamefont {Abbott}} \emph {et~al.} (\bibinfo {collaboration} {LIGO Scientific, Virgo, 1M2H, Dark Energy Camera GW-E, DES, DLT40, Las Cumbres Observatory, VINROUGE, MASTER}),\ }\bibfield  {title} {\enquote {\bibinfo {title} {{A gravitational-wave standard siren measurement of the Hubble constant}},}\ }\href {\doibase 10.1038/nature24471} {\bibfield  {journal} {\bibinfo  {journal} {Nature}\ }\textbf {\bibinfo {volume} {551}},\ \bibinfo {pages} {85--88} (\bibinfo {year} {2017}{\natexlab{a}})},\ \Eprint {http://arxiv.org/abs/1710.05835} {arXiv:1710.05835 [astro-ph.CO]} \BibitemShut {NoStop}%
\bibitem [{\citenamefont {Abbott}\ \emph {et~al.}(2021{\natexlab{c}})\citenamefont {Abbott} \emph {et~al.}}]{LIGOScientific:2019zcs}%
  \BibitemOpen
  \bibfield  {author} {\bibinfo {author} {\bibfnamefont {B.~P.}\ \bibnamefont {Abbott}} \emph {et~al.} (\bibinfo {collaboration} {LIGO Scientific, Virgo, VIRGO}),\ }\bibfield  {title} {\enquote {\bibinfo {title} {{A Gravitational-wave Measurement of the Hubble Constant Following the Second Observing Run of Advanced LIGO and Virgo}},}\ }\href {\doibase 10.3847/1538-4357/abdcb7} {\bibfield  {journal} {\bibinfo  {journal} {Astrophys. J.}\ }\textbf {\bibinfo {volume} {909}},\ \bibinfo {pages} {218} (\bibinfo {year} {2021}{\natexlab{c}})},\ \Eprint {http://arxiv.org/abs/1908.06060} {arXiv:1908.06060 [astro-ph.CO]} \BibitemShut {NoStop}%
\bibitem [{\citenamefont {Abbott}\ \emph {et~al.}(2023{\natexlab{c}})\citenamefont {Abbott} \emph {et~al.}}]{LIGOScientific:2021aug}%
  \BibitemOpen
  \bibfield  {author} {\bibinfo {author} {\bibfnamefont {R.}~\bibnamefont {Abbott}} \emph {et~al.} (\bibinfo {collaboration} {LIGO Scientific, Virgo, KAGRA}),\ }\bibfield  {title} {\enquote {\bibinfo {title} {{Constraints on the Cosmic Expansion History from GWTC\textendash{}3}},}\ }\href {\doibase 10.3847/1538-4357/ac74bb} {\bibfield  {journal} {\bibinfo  {journal} {Astrophys. J.}\ }\textbf {\bibinfo {volume} {949}},\ \bibinfo {pages} {76} (\bibinfo {year} {2023}{\natexlab{c}})},\ \Eprint {http://arxiv.org/abs/2111.03604} {arXiv:2111.03604 [astro-ph.CO]} \BibitemShut {NoStop}%
\bibitem [{\citenamefont {Schutz}(1986)}]{Schutz:1986gp}%
  \BibitemOpen
  \bibfield  {author} {\bibinfo {author} {\bibfnamefont {Bernard~F.}\ \bibnamefont {Schutz}},\ }\bibfield  {title} {\enquote {\bibinfo {title} {{Determining the Hubble Constant from Gravitational Wave Observations}},}\ }\href {\doibase 10.1038/323310a0} {\bibfield  {journal} {\bibinfo  {journal} {Nature}\ }\textbf {\bibinfo {volume} {323}},\ \bibinfo {pages} {310--311} (\bibinfo {year} {1986})}\BibitemShut {NoStop}%
\bibitem [{\citenamefont {Dalal}\ \emph {et~al.}(2006)\citenamefont {Dalal}, \citenamefont {Holz}, \citenamefont {Hughes},\ and\ \citenamefont {Jain}}]{Dalal:2006qt}%
  \BibitemOpen
  \bibfield  {author} {\bibinfo {author} {\bibfnamefont {Neal}\ \bibnamefont {Dalal}}, \bibinfo {author} {\bibfnamefont {Daniel~E.}\ \bibnamefont {Holz}}, \bibinfo {author} {\bibfnamefont {Scott~A.}\ \bibnamefont {Hughes}}, \ and\ \bibinfo {author} {\bibfnamefont {Bhuvnesh}\ \bibnamefont {Jain}},\ }\bibfield  {title} {\enquote {\bibinfo {title} {{Short grb and binary black hole standard sirens as a probe of dark energy}},}\ }\href {\doibase 10.1103/PhysRevD.74.063006} {\bibfield  {journal} {\bibinfo  {journal} {Phys. Rev. D}\ }\textbf {\bibinfo {volume} {74}},\ \bibinfo {pages} {063006} (\bibinfo {year} {2006})},\ \Eprint {http://arxiv.org/abs/astro-ph/0601275} {arXiv:astro-ph/0601275} \BibitemShut {NoStop}%
\bibitem [{\citenamefont {Nissanke}\ \emph {et~al.}(2013)\citenamefont {Nissanke}, \citenamefont {Holz}, \citenamefont {Dalal}, \citenamefont {Hughes}, \citenamefont {Sievers},\ and\ \citenamefont {Hirata}}]{Nissanke:2013fka}%
  \BibitemOpen
  \bibfield  {author} {\bibinfo {author} {\bibfnamefont {Samaya}\ \bibnamefont {Nissanke}}, \bibinfo {author} {\bibfnamefont {Daniel~E.}\ \bibnamefont {Holz}}, \bibinfo {author} {\bibfnamefont {Neal}\ \bibnamefont {Dalal}}, \bibinfo {author} {\bibfnamefont {Scott~A.}\ \bibnamefont {Hughes}}, \bibinfo {author} {\bibfnamefont {Jonathan~L.}\ \bibnamefont {Sievers}}, \ and\ \bibinfo {author} {\bibfnamefont {Christopher~M.}\ \bibnamefont {Hirata}},\ }\bibfield  {title} {\enquote {\bibinfo {title} {{Determining the Hubble constant from gravitational wave observations of merging compact binaries}},}\ }\href@noop {} {\  (\bibinfo {year} {2013})},\ \Eprint {http://arxiv.org/abs/1307.2638} {arXiv:1307.2638 [astro-ph.CO]} \BibitemShut {NoStop}%
\bibitem [{\citenamefont {Abbott}\ \emph {et~al.}(2019{\natexlab{b}})\citenamefont {Abbott} \emph {et~al.}}]{LIGOScientific:2018hze}%
  \BibitemOpen
  \bibfield  {author} {\bibinfo {author} {\bibfnamefont {B.~P.}\ \bibnamefont {Abbott}} \emph {et~al.} (\bibinfo {collaboration} {LIGO Scientific, Virgo}),\ }\bibfield  {title} {\enquote {\bibinfo {title} {{Properties of the binary neutron star merger GW170817}},}\ }\href {\doibase 10.1103/PhysRevX.9.011001} {\bibfield  {journal} {\bibinfo  {journal} {Phys. Rev. X}\ }\textbf {\bibinfo {volume} {9}},\ \bibinfo {pages} {011001} (\bibinfo {year} {2019}{\natexlab{b}})},\ \Eprint {http://arxiv.org/abs/1805.11579} {arXiv:1805.11579 [gr-qc]} \BibitemShut {NoStop}%
\bibitem [{\citenamefont {Palmese}\ \emph {et~al.}(2024)\citenamefont {Palmese}, \citenamefont {Kaur}, \citenamefont {Hajela}, \citenamefont {Margutti}, \citenamefont {McDowell},\ and\ \citenamefont {MacFadyen}}]{Palmese:2023beh}%
  \BibitemOpen
  \bibfield  {author} {\bibinfo {author} {\bibfnamefont {A.}~\bibnamefont {Palmese}}, \bibinfo {author} {\bibfnamefont {R.}~\bibnamefont {Kaur}}, \bibinfo {author} {\bibfnamefont {A.}~\bibnamefont {Hajela}}, \bibinfo {author} {\bibfnamefont {R.}~\bibnamefont {Margutti}}, \bibinfo {author} {\bibfnamefont {A.}~\bibnamefont {McDowell}}, \ and\ \bibinfo {author} {\bibfnamefont {A.}~\bibnamefont {MacFadyen}},\ }\bibfield  {title} {\enquote {\bibinfo {title} {{Standard siren measurement of the Hubble constant using GW170817 and the latest observations of the electromagnetic counterpart afterglow}},}\ }\href {\doibase 10.1103/PhysRevD.109.063508} {\bibfield  {journal} {\bibinfo  {journal} {Phys. Rev. D}\ }\textbf {\bibinfo {volume} {109}},\ \bibinfo {pages} {063508} (\bibinfo {year} {2024})},\ \Eprint {http://arxiv.org/abs/2305.19914} {arXiv:2305.19914 [astro-ph.CO]} \BibitemShut {NoStop}%
\bibitem [{\citenamefont {Chen}\ \emph {et~al.}(2018)\citenamefont {Chen}, \citenamefont {Fishbach},\ and\ \citenamefont {Holz}}]{Chen:2017rfc}%
  \BibitemOpen
  \bibfield  {author} {\bibinfo {author} {\bibfnamefont {Hsin-Yu}\ \bibnamefont {Chen}}, \bibinfo {author} {\bibfnamefont {Maya}\ \bibnamefont {Fishbach}}, \ and\ \bibinfo {author} {\bibfnamefont {Daniel~E.}\ \bibnamefont {Holz}},\ }\bibfield  {title} {\enquote {\bibinfo {title} {{A two per cent Hubble constant measurement from standard sirens within five years}},}\ }\href {\doibase 10.1038/s41586-018-0606-0} {\bibfield  {journal} {\bibinfo  {journal} {Nature}\ }\textbf {\bibinfo {volume} {562}},\ \bibinfo {pages} {545--547} (\bibinfo {year} {2018})},\ \Eprint {http://arxiv.org/abs/1712.06531} {arXiv:1712.06531 [astro-ph.CO]} \BibitemShut {NoStop}%
\bibitem [{\citenamefont {Di~Valentino}\ \emph {et~al.}(2021)\citenamefont {Di~Valentino}, \citenamefont {Mena}, \citenamefont {Pan}, \citenamefont {Visinelli}, \citenamefont {Yang}, \citenamefont {Melchiorri}, \citenamefont {Mota}, \citenamefont {Riess},\ and\ \citenamefont {Silk}}]{DiValentino:2021izs}%
  \BibitemOpen
  \bibfield  {author} {\bibinfo {author} {\bibfnamefont {Eleonora}\ \bibnamefont {Di~Valentino}}, \bibinfo {author} {\bibfnamefont {Olga}\ \bibnamefont {Mena}}, \bibinfo {author} {\bibfnamefont {Supriya}\ \bibnamefont {Pan}}, \bibinfo {author} {\bibfnamefont {Luca}\ \bibnamefont {Visinelli}}, \bibinfo {author} {\bibfnamefont {Weiqiang}\ \bibnamefont {Yang}}, \bibinfo {author} {\bibfnamefont {Alessandro}\ \bibnamefont {Melchiorri}}, \bibinfo {author} {\bibfnamefont {David~F.}\ \bibnamefont {Mota}}, \bibinfo {author} {\bibfnamefont {Adam~G.}\ \bibnamefont {Riess}}, \ and\ \bibinfo {author} {\bibfnamefont {Joseph}\ \bibnamefont {Silk}},\ }\bibfield  {title} {\enquote {\bibinfo {title} {{In the realm of the Hubble tension{\textemdash}a review of solutions}},}\ }\href {\doibase 10.1088/1361-6382/ac086d} {\bibfield  {journal} {\bibinfo  {journal} {Class. Quant. Grav.}\ }\textbf {\bibinfo {volume} {38}},\ \bibinfo {pages} {153001} (\bibinfo {year} {2021})},\ \Eprint {http://arxiv.org/abs/2103.01183} {arXiv:2103.01183 [astro-ph.CO]} \BibitemShut {NoStop}%
\bibitem [{\citenamefont {Verde}\ \emph {et~al.}(2024)\citenamefont {Verde}, \citenamefont {Sch{\"o}neberg},\ and\ \citenamefont {Gil-Mar{\'\i}n}}]{Verde:2023lmm}%
  \BibitemOpen
  \bibfield  {author} {\bibinfo {author} {\bibfnamefont {Licia}\ \bibnamefont {Verde}}, \bibinfo {author} {\bibfnamefont {Nils}\ \bibnamefont {Sch{\"o}neberg}}, \ and\ \bibinfo {author} {\bibfnamefont {H{\'e}ctor}\ \bibnamefont {Gil-Mar{\'\i}n}},\ }\bibfield  {title} {\enquote {\bibinfo {title} {{A Tale of Many H0}},}\ }\href {\doibase 10.1146/annurev-astro-052622-033813} {\bibfield  {journal} {\bibinfo  {journal} {Ann. Rev. Astron. Astrophys.}\ }\textbf {\bibinfo {volume} {62}},\ \bibinfo {pages} {287--331} (\bibinfo {year} {2024})},\ \Eprint {http://arxiv.org/abs/2311.13305} {arXiv:2311.13305 [astro-ph.CO]} \BibitemShut {NoStop}%
\bibitem [{\citenamefont {Abbott}\ \emph {et~al.}(2017{\natexlab{b}})\citenamefont {Abbott} \emph {et~al.}}]{LIGOScientific:2017ync}%
  \BibitemOpen
  \bibfield  {author} {\bibinfo {author} {\bibfnamefont {B.~P.}\ \bibnamefont {Abbott}} \emph {et~al.} (\bibinfo {collaboration} {LIGO Scientific, Virgo, Fermi GBM, INTEGRAL, IceCube, AstroSat Cadmium Zinc Telluride Imager Team, IPN, Insight-Hxmt, ANTARES, Swift, AGILE Team, 1M2H Team, Dark Energy Camera GW-EM, DES, DLT40, GRAWITA, Fermi-LAT, ATCA, ASKAP, Las Cumbres Observatory Group, OzGrav, DWF (Deeper Wider Faster Program), AST3, CAASTRO, VINROUGE, MASTER, J-GEM, GROWTH, JAGWAR, CaltechNRAO, TTU-NRAO, NuSTAR, Pan-STARRS, MAXI Team, TZAC Consortium, KU, Nordic Optical Telescope, ePESSTO, GROND, Texas Tech University, SALT Group, TOROS, BOOTES, MWA, CALET, IKI-GW Follow-up, H.E.S.S., LOFAR, LWA, HAWC, Pierre Auger, ALMA, Euro VLBI Team, Pi of Sky, Chandra Team at McGill University, DFN, ATLAS Telescopes, High Time Resolution Universe Survey, RIMAS, RATIR, SKA South Africa/MeerKAT}),\ }\bibfield  {title} {\enquote {\bibinfo {title} {{Multi-messenger Observations of a Binary Neutron Star Merger}},}\ }\href {\doibase 10.3847/2041-8213/aa91c9} {\bibfield  {journal} {\bibinfo  {journal} {Astrophys. J. Lett.}\ }\textbf {\bibinfo {volume} {848}},\ \bibinfo {pages} {L12} (\bibinfo {year} {2017}{\natexlab{b}})},\ \Eprint {http://arxiv.org/abs/1710.05833} {arXiv:1710.05833 [astro-ph.HE]} \BibitemShut {NoStop}%
\bibitem [{\citenamefont {Holz}\ and\ \citenamefont {Hughes}(2005)}]{Holz:2005df}%
  \BibitemOpen
  \bibfield  {author} {\bibinfo {author} {\bibfnamefont {Daniel~E.}\ \bibnamefont {Holz}}\ and\ \bibinfo {author} {\bibfnamefont {Scott~A.}\ \bibnamefont {Hughes}},\ }\bibfield  {title} {\enquote {\bibinfo {title} {{Using gravitational-wave standard sirens}},}\ }\href {\doibase 10.1086/431341} {\bibfield  {journal} {\bibinfo  {journal} {Astrophys. J.}\ }\textbf {\bibinfo {volume} {629}},\ \bibinfo {pages} {15--22} (\bibinfo {year} {2005})},\ \Eprint {http://arxiv.org/abs/astro-ph/0504616} {arXiv:astro-ph/0504616} \BibitemShut {NoStop}%
\bibitem [{\citenamefont {Del~Pozzo}(2012)}]{DelPozzo:2011vcw}%
  \BibitemOpen
  \bibfield  {author} {\bibinfo {author} {\bibfnamefont {Walter}\ \bibnamefont {Del~Pozzo}},\ }\bibfield  {title} {\enquote {\bibinfo {title} {{Inference of the cosmological parameters from gravitational waves: application to second generation interferometers}},}\ }\href {\doibase 10.1103/PhysRevD.86.043011} {\bibfield  {journal} {\bibinfo  {journal} {Phys. Rev. D}\ }\textbf {\bibinfo {volume} {86}},\ \bibinfo {pages} {043011} (\bibinfo {year} {2012})},\ \Eprint {http://arxiv.org/abs/1108.1317} {arXiv:1108.1317 [astro-ph.CO]} \BibitemShut {NoStop}%
\bibitem [{\citenamefont {Gray}\ \emph {et~al.}(2023)\citenamefont {Gray} \emph {et~al.}}]{Gray:2023wgj}%
  \BibitemOpen
  \bibfield  {author} {\bibinfo {author} {\bibfnamefont {Rachel}\ \bibnamefont {Gray}} \emph {et~al.},\ }\bibfield  {title} {\enquote {\bibinfo {title} {{Joint cosmological and gravitational-wave population inference using dark sirens and galaxy catalogues}},}\ }\href {\doibase 10.1088/1475-7516/2023/12/023} {\bibfield  {journal} {\bibinfo  {journal} {JCAP}\ }\textbf {\bibinfo {volume} {12}},\ \bibinfo {pages} {023} (\bibinfo {year} {2023})},\ \Eprint {http://arxiv.org/abs/2308.02281} {arXiv:2308.02281 [astro-ph.CO]} \BibitemShut {NoStop}%
\bibitem [{\citenamefont {Namikawa}\ \emph {et~al.}(2016)\citenamefont {Namikawa}, \citenamefont {Nishizawa},\ and\ \citenamefont {Taruya}}]{Namikawa:2015prh}%
  \BibitemOpen
  \bibfield  {author} {\bibinfo {author} {\bibfnamefont {Toshiya}\ \bibnamefont {Namikawa}}, \bibinfo {author} {\bibfnamefont {Atsushi}\ \bibnamefont {Nishizawa}}, \ and\ \bibinfo {author} {\bibfnamefont {Atsushi}\ \bibnamefont {Taruya}},\ }\bibfield  {title} {\enquote {\bibinfo {title} {{Anisotropies of gravitational-wave standard sirens as a new cosmological probe without redshift information}},}\ }\href {\doibase 10.1103/PhysRevLett.116.121302} {\bibfield  {journal} {\bibinfo  {journal} {Phys. Rev. Lett.}\ }\textbf {\bibinfo {volume} {116}},\ \bibinfo {pages} {121302} (\bibinfo {year} {2016})},\ \Eprint {http://arxiv.org/abs/1511.04638} {arXiv:1511.04638 [astro-ph.CO]} \BibitemShut {NoStop}%
\bibitem [{\citenamefont {Oguri}(2016)}]{Oguri:2016dgk}%
  \BibitemOpen
  \bibfield  {author} {\bibinfo {author} {\bibfnamefont {Masamune}\ \bibnamefont {Oguri}},\ }\bibfield  {title} {\enquote {\bibinfo {title} {{Measuring the distance-redshift relation with the cross-correlation of gravitational wave standard sirens and galaxies}},}\ }\href {\doibase 10.1103/PhysRevD.93.083511} {\bibfield  {journal} {\bibinfo  {journal} {Phys. Rev. D}\ }\textbf {\bibinfo {volume} {93}},\ \bibinfo {pages} {083511} (\bibinfo {year} {2016})},\ \Eprint {http://arxiv.org/abs/1603.02356} {arXiv:1603.02356 [astro-ph.CO]} \BibitemShut {NoStop}%
\bibitem [{\citenamefont {Mukherjee}\ \emph {et~al.}(2024)\citenamefont {Mukherjee}, \citenamefont {Krolewski}, \citenamefont {Wandelt},\ and\ \citenamefont {Silk}}]{Mukherjee:2022afz}%
  \BibitemOpen
  \bibfield  {author} {\bibinfo {author} {\bibfnamefont {Suvodip}\ \bibnamefont {Mukherjee}}, \bibinfo {author} {\bibfnamefont {Alex}\ \bibnamefont {Krolewski}}, \bibinfo {author} {\bibfnamefont {Benjamin~D.}\ \bibnamefont {Wandelt}}, \ and\ \bibinfo {author} {\bibfnamefont {Joseph}\ \bibnamefont {Silk}},\ }\bibfield  {title} {\enquote {\bibinfo {title} {{Cross-correlating dark sirens and galaxies: constraints on $H_0$ from GWTC-3 of LIGO-Virgo-KAGRA}},}\ }\href {\doibase 10.3847/1538-4357/ad7d90} {\bibfield  {journal} {\bibinfo  {journal} {Astrophys. J.}\ }\textbf {\bibinfo {volume} {975}},\ \bibinfo {pages} {189} (\bibinfo {year} {2024})},\ \Eprint {http://arxiv.org/abs/2203.03643} {arXiv:2203.03643 [astro-ph.CO]} \BibitemShut {NoStop}%
\bibitem [{\citenamefont {Mastrogiovanni}\ \emph {et~al.}(2023)\citenamefont {Mastrogiovanni}, \citenamefont {Laghi}, \citenamefont {Gray}, \citenamefont {Santoro}, \citenamefont {Ghosh}, \citenamefont {Karathanasis}, \citenamefont {Leyde}, \citenamefont {Steer}, \citenamefont {Perries},\ and\ \citenamefont {Pierra}}]{Mastrogiovanni:2023emh}%
  \BibitemOpen
  \bibfield  {author} {\bibinfo {author} {\bibfnamefont {Simone}\ \bibnamefont {Mastrogiovanni}}, \bibinfo {author} {\bibfnamefont {Danny}\ \bibnamefont {Laghi}}, \bibinfo {author} {\bibfnamefont {Rachel}\ \bibnamefont {Gray}}, \bibinfo {author} {\bibfnamefont {Giada~Caneva}\ \bibnamefont {Santoro}}, \bibinfo {author} {\bibfnamefont {Archisman}\ \bibnamefont {Ghosh}}, \bibinfo {author} {\bibfnamefont {Christos}\ \bibnamefont {Karathanasis}}, \bibinfo {author} {\bibfnamefont {Konstantin}\ \bibnamefont {Leyde}}, \bibinfo {author} {\bibfnamefont {Daniele~A.}\ \bibnamefont {Steer}}, \bibinfo {author} {\bibfnamefont {Stephane}\ \bibnamefont {Perries}}, \ and\ \bibinfo {author} {\bibfnamefont {Gregoire}\ \bibnamefont {Pierra}},\ }\bibfield  {title} {\enquote {\bibinfo {title} {{Joint population and cosmological properties inference with gravitational waves standard sirens and galaxy surveys}},}\ }\href {\doibase 10.1103/PhysRevD.108.042002} {\bibfield  {journal} {\bibinfo  {journal} {Phys. Rev. D}\ }\textbf {\bibinfo {volume} {108}},\ \bibinfo {pages} {042002} (\bibinfo {year} {2023})},\ \Eprint {http://arxiv.org/abs/2305.10488} {arXiv:2305.10488 [astro-ph.CO]} \BibitemShut {NoStop}%
\bibitem [{\citenamefont {Markovic}(1993)}]{Markovic:1993cr}%
  \BibitemOpen
  \bibfield  {author} {\bibinfo {author} {\bibfnamefont {Dragoljub}\ \bibnamefont {Markovic}},\ }\bibfield  {title} {\enquote {\bibinfo {title} {{On the possibility of determining cosmological parameters from measurements of gravitational waves emitted by coalescing, compact binaries}},}\ }\href {\doibase 10.1103/PhysRevD.48.4738} {\bibfield  {journal} {\bibinfo  {journal} {Phys. Rev. D}\ }\textbf {\bibinfo {volume} {48}},\ \bibinfo {pages} {4738--4756} (\bibinfo {year} {1993})}\BibitemShut {NoStop}%
\bibitem [{\citenamefont {Chernoff}\ and\ \citenamefont {Finn}(1993)}]{Chernoff:1993th}%
  \BibitemOpen
  \bibfield  {author} {\bibinfo {author} {\bibfnamefont {David~F.}\ \bibnamefont {Chernoff}}\ and\ \bibinfo {author} {\bibfnamefont {Lee~Samuel}\ \bibnamefont {Finn}},\ }\bibfield  {title} {\enquote {\bibinfo {title} {{Gravitational radiation, inspiraling binaries, and cosmology}},}\ }\href {\doibase 10.1086/186898} {\bibfield  {journal} {\bibinfo  {journal} {Astrophys. J. Lett.}\ }\textbf {\bibinfo {volume} {411}},\ \bibinfo {pages} {L5--L8} (\bibinfo {year} {1993})},\ \Eprint {http://arxiv.org/abs/gr-qc/9304020} {arXiv:gr-qc/9304020} \BibitemShut {NoStop}%
\bibitem [{\citenamefont {Finn}(1995)}]{Finn:1995wx}%
  \BibitemOpen
  \bibfield  {author} {\bibinfo {author} {\bibfnamefont {Lee~Samuel}\ \bibnamefont {Finn}},\ }\bibfield  {title} {\enquote {\bibinfo {title} {{Binary neutron star inspiral, LIGO, and cosmology}},}\ }\href {\doibase 10.1111/j.1749-6632.1995.tb17592.x} {\bibfield  {journal} {\bibinfo  {journal} {Annals N. Y. Acad. Sci.}\ }\textbf {\bibinfo {volume} {759}},\ \bibinfo {pages} {489} (\bibinfo {year} {1995})},\ \Eprint {http://arxiv.org/abs/gr-qc/9502013} {arXiv:gr-qc/9502013} \BibitemShut {NoStop}%
\bibitem [{\citenamefont {Taylor}\ \emph {et~al.}(2012)\citenamefont {Taylor}, \citenamefont {Gair},\ and\ \citenamefont {Mandel}}]{Taylor:2011fs}%
  \BibitemOpen
  \bibfield  {author} {\bibinfo {author} {\bibfnamefont {Stephen~R.}\ \bibnamefont {Taylor}}, \bibinfo {author} {\bibfnamefont {Jonathan~R.}\ \bibnamefont {Gair}}, \ and\ \bibinfo {author} {\bibfnamefont {Ilya}\ \bibnamefont {Mandel}},\ }\bibfield  {title} {\enquote {\bibinfo {title} {{Hubble without the Hubble: Cosmology using advanced gravitational-wave detectors alone}},}\ }\href {\doibase 10.1103/PhysRevD.85.023535} {\bibfield  {journal} {\bibinfo  {journal} {Phys. Rev. D}\ }\textbf {\bibinfo {volume} {85}},\ \bibinfo {pages} {023535} (\bibinfo {year} {2012})},\ \Eprint {http://arxiv.org/abs/1108.5161} {arXiv:1108.5161 [gr-qc]} \BibitemShut {NoStop}%
\bibitem [{\citenamefont {Farr}\ \emph {et~al.}(2019)\citenamefont {Farr}, \citenamefont {Fishbach}, \citenamefont {Ye},\ and\ \citenamefont {Holz}}]{Farr:2019twy}%
  \BibitemOpen
  \bibfield  {author} {\bibinfo {author} {\bibfnamefont {Will~M.}\ \bibnamefont {Farr}}, \bibinfo {author} {\bibfnamefont {Maya}\ \bibnamefont {Fishbach}}, \bibinfo {author} {\bibfnamefont {Jiani}\ \bibnamefont {Ye}}, \ and\ \bibinfo {author} {\bibfnamefont {Daniel}\ \bibnamefont {Holz}},\ }\bibfield  {title} {\enquote {\bibinfo {title} {{A Future Percent-Level Measurement of the Hubble Expansion at Redshift 0.8 With Advanced LIGO}},}\ }\href {\doibase 10.3847/2041-8213/ab4284} {\bibfield  {journal} {\bibinfo  {journal} {Astrophys. J. Lett.}\ }\textbf {\bibinfo {volume} {883}},\ \bibinfo {pages} {L42} (\bibinfo {year} {2019})},\ \Eprint {http://arxiv.org/abs/1908.09084} {arXiv:1908.09084 [astro-ph.CO]} \BibitemShut {NoStop}%
\bibitem [{\citenamefont {Ezquiaga}\ and\ \citenamefont {Holz}(2022)}]{Ezquiaga:2022zkx}%
  \BibitemOpen
  \bibfield  {author} {\bibinfo {author} {\bibfnamefont {Jose~Mar\'\i{}a}\ \bibnamefont {Ezquiaga}}\ and\ \bibinfo {author} {\bibfnamefont {Daniel~E.}\ \bibnamefont {Holz}},\ }\bibfield  {title} {\enquote {\bibinfo {title} {{Spectral Sirens: Cosmology from the Full Mass Distribution of Compact Binaries}},}\ }\href {\doibase 10.1103/PhysRevLett.129.061102} {\bibfield  {journal} {\bibinfo  {journal} {Phys. Rev. Lett.}\ }\textbf {\bibinfo {volume} {129}},\ \bibinfo {pages} {061102} (\bibinfo {year} {2022})},\ \Eprint {http://arxiv.org/abs/2202.08240} {arXiv:2202.08240 [astro-ph.CO]} \BibitemShut {NoStop}%
\bibitem [{\citenamefont {Mastrogiovanni}\ \emph {et~al.}(2022)\citenamefont {Mastrogiovanni}, \citenamefont {Leyde}, \citenamefont {Karathanasis}, \citenamefont {Chassande-Mottin}, \citenamefont {Steer}, \citenamefont {Gair}, \citenamefont {Ghosh}, \citenamefont {Gray}, \citenamefont {Mukherjee},\ and\ \citenamefont {Rinaldi}}]{Mastrogiovanni:2022hil}%
  \BibitemOpen
  \bibfield  {author} {\bibinfo {author} {\bibfnamefont {Simone}\ \bibnamefont {Mastrogiovanni}}, \bibinfo {author} {\bibfnamefont {Konstantin}\ \bibnamefont {Leyde}}, \bibinfo {author} {\bibfnamefont {Christos}\ \bibnamefont {Karathanasis}}, \bibinfo {author} {\bibfnamefont {Eric}\ \bibnamefont {Chassande-Mottin}}, \bibinfo {author} {\bibfnamefont {Daniele~Ann}\ \bibnamefont {Steer}}, \bibinfo {author} {\bibfnamefont {Jonathan}\ \bibnamefont {Gair}}, \bibinfo {author} {\bibfnamefont {Archisman}\ \bibnamefont {Ghosh}}, \bibinfo {author} {\bibfnamefont {Rachel}\ \bibnamefont {Gray}}, \bibinfo {author} {\bibfnamefont {Suvodip}\ \bibnamefont {Mukherjee}}, \ and\ \bibinfo {author} {\bibfnamefont {Stefano}\ \bibnamefont {Rinaldi}},\ }\bibfield  {title} {\enquote {\bibinfo {title} {{Cosmology in the dark: How compact binaries formation impact the gravitational-waves cosmological measurements}},}\ }\href {\doibase 10.22323/1.398.0098} {\bibfield  {journal} {\bibinfo  {journal} {PoS}\ }\textbf {\bibinfo {volume} {EPS-HEP2021}},\ \bibinfo {pages} {098} (\bibinfo {year} {2022})},\ \Eprint {http://arxiv.org/abs/2205.05421} {arXiv:2205.05421 [gr-qc]} \BibitemShut {NoStop}%
\bibitem [{\citenamefont {Mali}\ and\ \citenamefont {Essick}(2024)}]{Mali:2024wpq}%
  \BibitemOpen
  \bibfield  {author} {\bibinfo {author} {\bibfnamefont {Utkarsh}\ \bibnamefont {Mali}}\ and\ \bibinfo {author} {\bibfnamefont {Reed}\ \bibnamefont {Essick}},\ }\bibfield  {title} {\enquote {\bibinfo {title} {{Striking a Chord with Spectral Sirens: multiple features in the compact binary population correlate with $H_0$}},}\ }\href@noop {} {\  (\bibinfo {year} {2024})},\ \Eprint {http://arxiv.org/abs/2410.07416} {arXiv:2410.07416 [astro-ph.HE]} \BibitemShut {NoStop}%
\bibitem [{\citenamefont {Abbott}\ \emph {et~al.}(2016{\natexlab{b}})\citenamefont {Abbott} \emph {et~al.}}]{KAGRA:2013rdx}%
  \BibitemOpen
  \bibfield  {author} {\bibinfo {author} {\bibfnamefont {B.~P.}\ \bibnamefont {Abbott}} \emph {et~al.} (\bibinfo {collaboration} {KAGRA, LIGO Scientific, Virgo}),\ }\bibfield  {title} {\enquote {\bibinfo {title} {{Prospects for observing and localizing gravitational-wave transients with Advanced LIGO, Advanced Virgo and KAGRA}},}\ }\href {\doibase 10.1007/s41114-020-00026-9} {\bibfield  {journal} {\bibinfo  {journal} {Living Rev. Rel.}\ }\textbf {\bibinfo {volume} {19}},\ \bibinfo {pages} {1} (\bibinfo {year} {2016}{\natexlab{b}})},\ \Eprint {http://arxiv.org/abs/1304.0670} {arXiv:1304.0670 [gr-qc]} \BibitemShut {NoStop}%
\bibitem [{\citenamefont {Nicolaou}\ \emph {et~al.}(2020)\citenamefont {Nicolaou}, \citenamefont {Lahav}, \citenamefont {Lemos}, \citenamefont {Hartley},\ and\ \citenamefont {Braden}}]{Nicolaou:2019cip}%
  \BibitemOpen
  \bibfield  {author} {\bibinfo {author} {\bibfnamefont {Constantina}\ \bibnamefont {Nicolaou}}, \bibinfo {author} {\bibfnamefont {Ofer}\ \bibnamefont {Lahav}}, \bibinfo {author} {\bibfnamefont {Pablo}\ \bibnamefont {Lemos}}, \bibinfo {author} {\bibfnamefont {William}\ \bibnamefont {Hartley}}, \ and\ \bibinfo {author} {\bibfnamefont {Jonathan}\ \bibnamefont {Braden}},\ }\bibfield  {title} {\enquote {\bibinfo {title} {{The Impact of Peculiar Velocities on the Estimation of the Hubble Constant from Gravitational Wave Standard Sirens}},}\ }\href {\doibase 10.1093/mnras/staa1120} {\bibfield  {journal} {\bibinfo  {journal} {Mon. Not. Roy. Astron. Soc.}\ }\textbf {\bibinfo {volume} {495}},\ \bibinfo {pages} {90--97} (\bibinfo {year} {2020})},\ \Eprint {http://arxiv.org/abs/1909.09609} {arXiv:1909.09609 [astro-ph.CO]} \BibitemShut {NoStop}%
\bibitem [{\citenamefont {Hotokezaka}\ \emph {et~al.}(2019)\citenamefont {Hotokezaka}, \citenamefont {Nakar}, \citenamefont {Gottlieb}, \citenamefont {Nissanke}, \citenamefont {Masuda}, \citenamefont {Hallinan}, \citenamefont {Mooley},\ and\ \citenamefont {Deller}}]{Hotokezaka:2018dfi}%
  \BibitemOpen
  \bibfield  {author} {\bibinfo {author} {\bibfnamefont {Kenta}\ \bibnamefont {Hotokezaka}}, \bibinfo {author} {\bibfnamefont {Ehud}\ \bibnamefont {Nakar}}, \bibinfo {author} {\bibfnamefont {Ore}\ \bibnamefont {Gottlieb}}, \bibinfo {author} {\bibfnamefont {Samaya}\ \bibnamefont {Nissanke}}, \bibinfo {author} {\bibfnamefont {Kento}\ \bibnamefont {Masuda}}, \bibinfo {author} {\bibfnamefont {Gregg}\ \bibnamefont {Hallinan}}, \bibinfo {author} {\bibfnamefont {Kunal~P.}\ \bibnamefont {Mooley}}, \ and\ \bibinfo {author} {\bibfnamefont {Adam.~T.}\ \bibnamefont {Deller}},\ }\bibfield  {title} {\enquote {\bibinfo {title} {{A Hubble constant measurement from superluminal motion of the jet in GW170817}},}\ }\href {\doibase 10.1038/s41550-019-0820-1} {\bibfield  {journal} {\bibinfo  {journal} {Nature Astron.}\ }\textbf {\bibinfo {volume} {3}},\ \bibinfo {pages} {940--944} (\bibinfo {year} {2019})},\ \Eprint {http://arxiv.org/abs/1806.10596} {arXiv:1806.10596 [astro-ph.CO]} \BibitemShut {NoStop}%
\bibitem [{\citenamefont {Mukherjee}\ \emph {et~al.}(2021)\citenamefont {Mukherjee}, \citenamefont {Lavaux}, \citenamefont {Bouchet}, \citenamefont {Jasche}, \citenamefont {Wandelt}, \citenamefont {Nissanke}, \citenamefont {Leclercq},\ and\ \citenamefont {Hotokezaka}}]{Mukherjee:2019qmm}%
  \BibitemOpen
  \bibfield  {author} {\bibinfo {author} {\bibfnamefont {Suvodip}\ \bibnamefont {Mukherjee}}, \bibinfo {author} {\bibfnamefont {Guilhem}\ \bibnamefont {Lavaux}}, \bibinfo {author} {\bibfnamefont {Fran{\c{c}}ois~R.}\ \bibnamefont {Bouchet}}, \bibinfo {author} {\bibfnamefont {Jens}\ \bibnamefont {Jasche}}, \bibinfo {author} {\bibfnamefont {Benjamin~D.}\ \bibnamefont {Wandelt}}, \bibinfo {author} {\bibfnamefont {Samaya~M.}\ \bibnamefont {Nissanke}}, \bibinfo {author} {\bibfnamefont {Florent}\ \bibnamefont {Leclercq}}, \ and\ \bibinfo {author} {\bibfnamefont {Kenta}\ \bibnamefont {Hotokezaka}},\ }\bibfield  {title} {\enquote {\bibinfo {title} {{Velocity correction for Hubble constant measurements from standard sirens}},}\ }\href {\doibase 10.1051/0004-6361/201936724} {\bibfield  {journal} {\bibinfo  {journal} {Astron. Astrophys.}\ }\textbf {\bibinfo {volume} {646}},\ \bibinfo {pages} {A65} (\bibinfo {year} {2021})},\ \Eprint {http://arxiv.org/abs/1909.08627} {arXiv:1909.08627 [astro-ph.CO]} \BibitemShut {NoStop}%
\bibitem [{\citenamefont {Abac}\ \emph {et~al.}(2025{\natexlab{a}})\citenamefont {Abac} \emph {et~al.}}]{LIGOScientific:2025hdt}%
  \BibitemOpen
  \bibfield  {author} {\bibinfo {author} {\bibfnamefont {A.~G.}\ \bibnamefont {Abac}} \emph {et~al.} (\bibinfo {collaboration} {LIGO Scientific, VIRGO, KAGRA}),\ }\bibfield  {title} {\enquote {\bibinfo {title} {{GWTC-4.0: An Introduction to Version 4.0 of the Gravitational-Wave Transient Catalog}},}\ }\href@noop {} {\  (\bibinfo {year} {2025}{\natexlab{a}})},\ \Eprint {http://arxiv.org/abs/2508.18080} {arXiv:2508.18080 [gr-qc]} \BibitemShut {NoStop}%
\bibitem [{\citenamefont {Abac}\ \emph {et~al.}(2025{\natexlab{b}})\citenamefont {Abac} \emph {et~al.}}]{LIGOScientific:2025slb}%
  \BibitemOpen
  \bibfield  {author} {\bibinfo {author} {\bibfnamefont {A.~G.}\ \bibnamefont {Abac}} \emph {et~al.} (\bibinfo {collaboration} {LIGO Scientific, VIRGO, KAGRA}),\ }\bibfield  {title} {\enquote {\bibinfo {title} {{GWTC-4.0: Updating the Gravitational-Wave Transient Catalog with Observations from the First Part of the Fourth LIGO-Virgo-KAGRA Observing Run}},}\ }\href@noop {} {\  (\bibinfo {year} {2025}{\natexlab{b}})},\ \Eprint {http://arxiv.org/abs/2508.18082} {arXiv:2508.18082 [gr-qc]} \BibitemShut {NoStop}%
\bibitem [{\citenamefont {Abac}\ \emph {et~al.}(2025{\natexlab{c}})\citenamefont {Abac} \emph {et~al.}}]{LIGOScientific:2025yae}%
  \BibitemOpen
  \bibfield  {author} {\bibinfo {author} {\bibfnamefont {A.~G.}\ \bibnamefont {Abac}} \emph {et~al.} (\bibinfo {collaboration} {LIGO Scientific, VIRGO, KAGRA}),\ }\bibfield  {title} {\enquote {\bibinfo {title} {{GWTC-4.0: Methods for Identifying and Characterizing Gravitational-wave Transients}},}\ }\href@noop {} {\  (\bibinfo {year} {2025}{\natexlab{c}})},\ \Eprint {http://arxiv.org/abs/2508.18081} {arXiv:2508.18081 [gr-qc]} \BibitemShut {NoStop}%
\bibitem [{LIG(2025)}]{LIGOScientific:2025jau}%
  \BibitemOpen
  \bibfield  {title} {\enquote {\bibinfo {title} {{GWTC-4.0: Constraints on the Cosmic Expansion Rate and Modified Gravitational-wave Propagation}},}\ }\href@noop {} {\  (\bibinfo {year} {2025})},\ \Eprint {http://arxiv.org/abs/2509.04348} {arXiv:2509.04348 [astro-ph.CO]} \BibitemShut {NoStop}%
\bibitem [{\citenamefont {Tiwari}(2022)}]{Tiwari:2021yvr}%
  \BibitemOpen
  \bibfield  {author} {\bibinfo {author} {\bibfnamefont {Vaibhav}\ \bibnamefont {Tiwari}},\ }\bibfield  {title} {\enquote {\bibinfo {title} {{Exploring Features in the Binary Black Hole Population}},}\ }\href {\doibase 10.3847/1538-4357/ac589a} {\bibfield  {journal} {\bibinfo  {journal} {Astrophys. J.}\ }\textbf {\bibinfo {volume} {928}},\ \bibinfo {pages} {155} (\bibinfo {year} {2022})},\ \Eprint {http://arxiv.org/abs/2111.13991} {arXiv:2111.13991 [astro-ph.HE]} \BibitemShut {NoStop}%
\bibitem [{\citenamefont {Sadiq}\ \emph {et~al.}(2022)\citenamefont {Sadiq}, \citenamefont {Dent},\ and\ \citenamefont {Wysocki}}]{Sadiq:2021fin}%
  \BibitemOpen
  \bibfield  {author} {\bibinfo {author} {\bibfnamefont {Jam}\ \bibnamefont {Sadiq}}, \bibinfo {author} {\bibfnamefont {Thomas}\ \bibnamefont {Dent}}, \ and\ \bibinfo {author} {\bibfnamefont {Daniel}\ \bibnamefont {Wysocki}},\ }\bibfield  {title} {\enquote {\bibinfo {title} {{Flexible and fast estimation of binary merger population distributions with an adaptive kernel density estimator}},}\ }\href {\doibase 10.1103/PhysRevD.105.123014} {\bibfield  {journal} {\bibinfo  {journal} {Phys. Rev. D}\ }\textbf {\bibinfo {volume} {105}},\ \bibinfo {pages} {123014} (\bibinfo {year} {2022})},\ \Eprint {http://arxiv.org/abs/2112.12659} {arXiv:2112.12659 [gr-qc]} \BibitemShut {NoStop}%
\bibitem [{\citenamefont {Farah}\ \emph {et~al.}(2023)\citenamefont {Farah}, \citenamefont {Edelman}, \citenamefont {Zevin}, \citenamefont {Fishbach}, \citenamefont {Ezquiaga}, \citenamefont {Farr},\ and\ \citenamefont {Holz}}]{Farah:2023vsc}%
  \BibitemOpen
  \bibfield  {author} {\bibinfo {author} {\bibfnamefont {Amanda~M.}\ \bibnamefont {Farah}}, \bibinfo {author} {\bibfnamefont {Bruce}\ \bibnamefont {Edelman}}, \bibinfo {author} {\bibfnamefont {Michael}\ \bibnamefont {Zevin}}, \bibinfo {author} {\bibfnamefont {Maya}\ \bibnamefont {Fishbach}}, \bibinfo {author} {\bibfnamefont {Jose~Mar\'\i{}a}\ \bibnamefont {Ezquiaga}}, \bibinfo {author} {\bibfnamefont {Ben}\ \bibnamefont {Farr}}, \ and\ \bibinfo {author} {\bibfnamefont {Daniel~E.}\ \bibnamefont {Holz}},\ }\bibfield  {title} {\enquote {\bibinfo {title} {{Things That Might Go Bump in the Night: Assessing Structure in the Binary Black Hole Mass Spectrum}},}\ }\href {\doibase 10.3847/1538-4357/aced02} {\bibfield  {journal} {\bibinfo  {journal} {Astrophys. J.}\ }\textbf {\bibinfo {volume} {955}},\ \bibinfo {pages} {107} (\bibinfo {year} {2023})},\ \Eprint {http://arxiv.org/abs/2301.00834} {arXiv:2301.00834 [astro-ph.HE]} \BibitemShut {NoStop}%
\bibitem [{\citenamefont {Toubiana}\ \emph {et~al.}(2023)\citenamefont {Toubiana}, \citenamefont {Katz},\ and\ \citenamefont {Gair}}]{Toubiana:2023egi}%
  \BibitemOpen
  \bibfield  {author} {\bibinfo {author} {\bibfnamefont {Alexandre}\ \bibnamefont {Toubiana}}, \bibinfo {author} {\bibfnamefont {Michael~L.}\ \bibnamefont {Katz}}, \ and\ \bibinfo {author} {\bibfnamefont {Jonathan~R.}\ \bibnamefont {Gair}},\ }\bibfield  {title} {\enquote {\bibinfo {title} {{Is there an excess of black holes around 20 M\ensuremath{\odot}? Optimizing the complexity of population models with the use of reversible jump MCMC.}}}\ }\href {\doibase 10.1093/mnras/stad2215} {\bibfield  {journal} {\bibinfo  {journal} {Mon. Not. Roy. Astron. Soc.}\ }\textbf {\bibinfo {volume} {524}},\ \bibinfo {pages} {5844--5853} (\bibinfo {year} {2023})},\ \Eprint {http://arxiv.org/abs/2305.08909} {arXiv:2305.08909 [gr-qc]} \BibitemShut {NoStop}%
\bibitem [{\citenamefont {Gennari}\ \emph {et~al.}(2025)\citenamefont {Gennari}, \citenamefont {Mastrogiovanni}, \citenamefont {Tamanini}, \citenamefont {Marsat},\ and\ \citenamefont {Pierra}}]{Gennari:2025nho}%
  \BibitemOpen
  \bibfield  {author} {\bibinfo {author} {\bibfnamefont {Vasco}\ \bibnamefont {Gennari}}, \bibinfo {author} {\bibfnamefont {Simone}\ \bibnamefont {Mastrogiovanni}}, \bibinfo {author} {\bibfnamefont {Nicola}\ \bibnamefont {Tamanini}}, \bibinfo {author} {\bibfnamefont {Sylvain}\ \bibnamefont {Marsat}}, \ and\ \bibinfo {author} {\bibfnamefont {Gr{\'e}goire}\ \bibnamefont {Pierra}},\ }\bibfield  {title} {\enquote {\bibinfo {title} {{Searching for additional structure and redshift evolution in the observed binary black hole population with a parametric time-dependent mass distribution}},}\ }\href {\doibase 10.1103/ftw9-7xd5} {\bibfield  {journal} {\bibinfo  {journal} {Phys. Rev. D}\ }\textbf {\bibinfo {volume} {111}},\ \bibinfo {pages} {123046} (\bibinfo {year} {2025})},\ \Eprint {http://arxiv.org/abs/2502.20445} {arXiv:2502.20445 [gr-qc]} \BibitemShut {NoStop}%
\bibitem [{\citenamefont {Abac}\ \emph {et~al.}(2025{\natexlab{d}})\citenamefont {Abac} \emph {et~al.}}]{LIGOScientific:2025pvj}%
  \BibitemOpen
  \bibfield  {author} {\bibinfo {author} {\bibfnamefont {A.~G.}\ \bibnamefont {Abac}} \emph {et~al.} (\bibinfo {collaboration} {LIGO Scientific, VIRGO, KAGRA}),\ }\bibfield  {title} {\enquote {\bibinfo {title} {{GWTC-4.0: Population Properties of Merging Compact Binaries}},}\ }\href@noop {} {\  (\bibinfo {year} {2025}{\natexlab{d}})},\ \Eprint {http://arxiv.org/abs/2508.18083} {arXiv:2508.18083 [astro-ph.HE]} \BibitemShut {NoStop}%
\bibitem [{\citenamefont {Maga{\~n}a~Hernandez}\ and\ \citenamefont {Palmese}(2025)}]{MaganaHernandez:2025cnu}%
  \BibitemOpen
  \bibfield  {author} {\bibinfo {author} {\bibfnamefont {Ignacio}\ \bibnamefont {Maga{\~n}a~Hernandez}}\ and\ \bibinfo {author} {\bibfnamefont {Antonella}\ \bibnamefont {Palmese}},\ }\bibfield  {title} {\enquote {\bibinfo {title} {{Spectral siren cosmology from gravitational-wave observations in GWTC-4.0}},}\ }\href@noop {} {\  (\bibinfo {year} {2025})},\ \Eprint {http://arxiv.org/abs/2509.03607} {arXiv:2509.03607 [astro-ph.CO]} \BibitemShut {NoStop}%
\bibitem [{\citenamefont {Collaboration}\ \emph {et~al.}(2021)\citenamefont {Collaboration}, \citenamefont {Collaboration},\ and\ \citenamefont {Collaboration}}]{gwtc3:cosmo_release}%
  \BibitemOpen
  \bibfield  {author} {\bibinfo {author} {\bibfnamefont {LIGO~Scientific}\ \bibnamefont {Collaboration}}, \bibinfo {author} {\bibfnamefont {Virgo}\ \bibnamefont {Collaboration}}, \ and\ \bibinfo {author} {\bibfnamefont {KAGRA}\ \bibnamefont {Collaboration}},\ }\href@noop {} {\enquote {\bibinfo {title} {Data distribution of constraints on the cosmic expansion history from the gwtc-3 [data set]},}\ }\bibinfo {howpublished} {\url{https://doi.org/10.5281/zenodo.5645777}} (\bibinfo {year} {2021})\BibitemShut {NoStop}%
\bibitem [{\citenamefont {Gray}\ \emph {et~al.}(2020)\citenamefont {Gray} \emph {et~al.}}]{Gray:2019ksv}%
  \BibitemOpen
  \bibfield  {author} {\bibinfo {author} {\bibfnamefont {Rachel}\ \bibnamefont {Gray}} \emph {et~al.},\ }\bibfield  {title} {\enquote {\bibinfo {title} {{Cosmological inference using gravitational wave standard sirens: A mock data analysis}},}\ }\href {\doibase 10.1103/PhysRevD.101.122001} {\bibfield  {journal} {\bibinfo  {journal} {Phys. Rev. D}\ }\textbf {\bibinfo {volume} {101}},\ \bibinfo {pages} {122001} (\bibinfo {year} {2020})},\ \Eprint {http://arxiv.org/abs/1908.06050} {arXiv:1908.06050 [gr-qc]} \BibitemShut {NoStop}%
\bibitem [{\citenamefont {Gray}\ \emph {et~al.}(2022)\citenamefont {Gray}, \citenamefont {Messenger},\ and\ \citenamefont {Veitch}}]{Gray:2021sew}%
  \BibitemOpen
  \bibfield  {author} {\bibinfo {author} {\bibfnamefont {Rachel}\ \bibnamefont {Gray}}, \bibinfo {author} {\bibfnamefont {Chris}\ \bibnamefont {Messenger}}, \ and\ \bibinfo {author} {\bibfnamefont {John}\ \bibnamefont {Veitch}},\ }\bibfield  {title} {\enquote {\bibinfo {title} {{A pixelated approach to galaxy catalogue incompleteness: improving the dark siren measurement of the Hubble constant}},}\ }\href {\doibase 10.1093/mnras/stac366} {\bibfield  {journal} {\bibinfo  {journal} {Mon. Not. Roy. Astron. Soc.}\ }\textbf {\bibinfo {volume} {512}},\ \bibinfo {pages} {1127--1140} (\bibinfo {year} {2022})},\ \Eprint {http://arxiv.org/abs/2111.04629} {arXiv:2111.04629 [astro-ph.CO]} \BibitemShut {NoStop}%
\bibitem [{\citenamefont {Aghanim}\ \emph {et~al.}(2020)\citenamefont {Aghanim} \emph {et~al.}}]{Planck:2018vyg}%
  \BibitemOpen
  \bibfield  {author} {\bibinfo {author} {\bibfnamefont {N.}~\bibnamefont {Aghanim}} \emph {et~al.} (\bibinfo {collaboration} {Planck}),\ }\bibfield  {title} {\enquote {\bibinfo {title} {{Planck 2018 results. VI. Cosmological parameters}},}\ }\href {\doibase 10.1051/0004-6361/201833910} {\bibfield  {journal} {\bibinfo  {journal} {Astron. Astrophys.}\ }\textbf {\bibinfo {volume} {641}},\ \bibinfo {pages} {A6} (\bibinfo {year} {2020})},\ \bibinfo {note} {[Erratum: Astron.Astrophys. 652, C4 (2021)]},\ \Eprint {http://arxiv.org/abs/1807.06209} {arXiv:1807.06209 [astro-ph.CO]} \BibitemShut {NoStop}%
\bibitem [{\citenamefont {Riess}\ \emph {et~al.}(2022)\citenamefont {Riess} \emph {et~al.}}]{Riess:2021jrx}%
  \BibitemOpen
  \bibfield  {author} {\bibinfo {author} {\bibfnamefont {Adam~G.}\ \bibnamefont {Riess}} \emph {et~al.},\ }\bibfield  {title} {\enquote {\bibinfo {title} {{A Comprehensive Measurement of the Local Value of the Hubble Constant with 1 km s$^{-1}$ Mpc$^{-1}$ Uncertainty from the Hubble Space Telescope and the SH0ES Team}},}\ }\href {\doibase 10.3847/2041-8213/ac5c5b} {\bibfield  {journal} {\bibinfo  {journal} {Astrophys. J. Lett.}\ }\textbf {\bibinfo {volume} {934}},\ \bibinfo {pages} {L7} (\bibinfo {year} {2022})},\ \Eprint {http://arxiv.org/abs/2112.04510} {arXiv:2112.04510 [astro-ph.CO]} \BibitemShut {NoStop}%
\bibitem [{\citenamefont {Mastrogiovanni}\ \emph {et~al.}(2024)\citenamefont {Mastrogiovanni}, \citenamefont {Pierra}, \citenamefont {Perri\`es}, \citenamefont {Laghi}, \citenamefont {Caneva~Santoro}, \citenamefont {Ghosh}, \citenamefont {Gray}, \citenamefont {Karathanasis},\ and\ \citenamefont {Leyde}}]{Mastrogiovanni:2023zbw}%
  \BibitemOpen
  \bibfield  {author} {\bibinfo {author} {\bibfnamefont {Simone}\ \bibnamefont {Mastrogiovanni}}, \bibinfo {author} {\bibfnamefont {Gr\'egoire}\ \bibnamefont {Pierra}}, \bibinfo {author} {\bibfnamefont {St\'ephane}\ \bibnamefont {Perri\`es}}, \bibinfo {author} {\bibfnamefont {Danny}\ \bibnamefont {Laghi}}, \bibinfo {author} {\bibfnamefont {Giada}\ \bibnamefont {Caneva~Santoro}}, \bibinfo {author} {\bibfnamefont {Archisman}\ \bibnamefont {Ghosh}}, \bibinfo {author} {\bibfnamefont {Rachel}\ \bibnamefont {Gray}}, \bibinfo {author} {\bibfnamefont {Christos}\ \bibnamefont {Karathanasis}}, \ and\ \bibinfo {author} {\bibfnamefont {Konstantin}\ \bibnamefont {Leyde}},\ }\bibfield  {title} {\enquote {\bibinfo {title} {{ICAROGW: A python package for inference of astrophysical population properties of noisy, heterogeneous, and incomplete observations}},}\ }\href {\doibase 10.1051/0004-6361/202347007} {\bibfield  {journal} {\bibinfo  {journal} {Astron. Astrophys.}\ }\textbf {\bibinfo {volume} {682}},\ \bibinfo {pages} {A167} (\bibinfo {year} {2024})},\ \Eprint {http://arxiv.org/abs/2305.17973} {arXiv:2305.17973 [astro-ph.CO]} \BibitemShut {NoStop}%
\bibitem [{\citenamefont {Madau}\ and\ \citenamefont {Dickinson}(2014)}]{Madau:2014bja}%
  \BibitemOpen
  \bibfield  {author} {\bibinfo {author} {\bibfnamefont {Piero}\ \bibnamefont {Madau}}\ and\ \bibinfo {author} {\bibfnamefont {Mark}\ \bibnamefont {Dickinson}},\ }\bibfield  {title} {\enquote {\bibinfo {title} {{Cosmic Star Formation History}},}\ }\href {\doibase 10.1146/annurev-astro-081811-125615} {\bibfield  {journal} {\bibinfo  {journal} {Ann. Rev. Astron. Astrophys.}\ }\textbf {\bibinfo {volume} {52}},\ \bibinfo {pages} {415--486} (\bibinfo {year} {2014})},\ \Eprint {http://arxiv.org/abs/1403.0007} {arXiv:1403.0007 [astro-ph.CO]} \BibitemShut {NoStop}%
\bibitem [{\citenamefont {Aasi}\ \emph {et~al.}(2015)\citenamefont {Aasi} \emph {et~al.}}]{LIGOScientific:2014pky}%
  \BibitemOpen
  \bibfield  {author} {\bibinfo {author} {\bibfnamefont {J.}~\bibnamefont {Aasi}} \emph {et~al.} (\bibinfo {collaboration} {LIGO Scientific}),\ }\bibfield  {title} {\enquote {\bibinfo {title} {{Advanced LIGO}},}\ }\href {\doibase 10.1088/0264-9381/32/7/074001} {\bibfield  {journal} {\bibinfo  {journal} {Class. Quant. Grav.}\ }\textbf {\bibinfo {volume} {32}},\ \bibinfo {pages} {074001} (\bibinfo {year} {2015})},\ \Eprint {http://arxiv.org/abs/1411.4547} {arXiv:1411.4547 [gr-qc]} \BibitemShut {NoStop}%
\bibitem [{\citenamefont {Acernese}\ \emph {et~al.}(2015)\citenamefont {Acernese} \emph {et~al.}}]{VIRGO:2014yos}%
  \BibitemOpen
  \bibfield  {author} {\bibinfo {author} {\bibfnamefont {F.}~\bibnamefont {Acernese}} \emph {et~al.} (\bibinfo {collaboration} {VIRGO}),\ }\bibfield  {title} {\enquote {\bibinfo {title} {{Advanced Virgo: a second-generation interferometric gravitational wave detector}},}\ }\href {\doibase 10.1088/0264-9381/32/2/024001} {\bibfield  {journal} {\bibinfo  {journal} {Class. Quant. Grav.}\ }\textbf {\bibinfo {volume} {32}},\ \bibinfo {pages} {024001} (\bibinfo {year} {2015})},\ \Eprint {http://arxiv.org/abs/1408.3978} {arXiv:1408.3978 [gr-qc]} \BibitemShut {NoStop}%
\bibitem [{\citenamefont {Akutsu}\ \emph {et~al.}(2021)\citenamefont {Akutsu} \emph {et~al.}}]{KAGRA:2020tym}%
  \BibitemOpen
  \bibfield  {author} {\bibinfo {author} {\bibfnamefont {T.}~\bibnamefont {Akutsu}} \emph {et~al.} (\bibinfo {collaboration} {KAGRA}),\ }\bibfield  {title} {\enquote {\bibinfo {title} {{Overview of KAGRA: Detector design and construction history}},}\ }\href {\doibase 10.1093/ptep/ptaa125} {\bibfield  {journal} {\bibinfo  {journal} {PTEP}\ }\textbf {\bibinfo {volume} {2021}},\ \bibinfo {pages} {05A101} (\bibinfo {year} {2021})},\ \Eprint {http://arxiv.org/abs/2005.05574} {arXiv:2005.05574 [physics.ins-det]} \BibitemShut {NoStop}%
\bibitem [{\citenamefont {Ade}\ \emph {et~al.}(2016)\citenamefont {Ade} \emph {et~al.}}]{Planck:2015fie}%
  \BibitemOpen
  \bibfield  {author} {\bibinfo {author} {\bibfnamefont {P.~A.~R.}\ \bibnamefont {Ade}} \emph {et~al.} (\bibinfo {collaboration} {Planck}),\ }\bibfield  {title} {\enquote {\bibinfo {title} {{Planck 2015 results. XIII. Cosmological parameters}},}\ }\href {\doibase 10.1051/0004-6361/201525830} {\bibfield  {journal} {\bibinfo  {journal} {Astron. Astrophys.}\ }\textbf {\bibinfo {volume} {594}},\ \bibinfo {pages} {A13} (\bibinfo {year} {2016})},\ \Eprint {http://arxiv.org/abs/1502.01589} {arXiv:1502.01589 [astro-ph.CO]} \BibitemShut {NoStop}%
\bibitem [{\citenamefont {Ashton}\ \emph {et~al.}(2019)\citenamefont {Ashton} \emph {et~al.}}]{Ashton:2018jfp}%
  \BibitemOpen
  \bibfield  {author} {\bibinfo {author} {\bibfnamefont {Gregory}\ \bibnamefont {Ashton}} \emph {et~al.},\ }\bibfield  {title} {\enquote {\bibinfo {title} {{BILBY: A user-friendly Bayesian inference library for gravitational-wave astronomy}},}\ }\href {\doibase 10.3847/1538-4365/ab06fc} {\bibfield  {journal} {\bibinfo  {journal} {Astrophys. J. Suppl.}\ }\textbf {\bibinfo {volume} {241}},\ \bibinfo {pages} {27} (\bibinfo {year} {2019})},\ \Eprint {http://arxiv.org/abs/1811.02042} {arXiv:1811.02042 [astro-ph.IM]} \BibitemShut {NoStop}%
\bibitem [{\citenamefont {Romero-Shaw}\ \emph {et~al.}(2020)\citenamefont {Romero-Shaw} \emph {et~al.}}]{Romero-Shaw:2020owr}%
  \BibitemOpen
  \bibfield  {author} {\bibinfo {author} {\bibfnamefont {I.~M.}\ \bibnamefont {Romero-Shaw}} \emph {et~al.},\ }\bibfield  {title} {\enquote {\bibinfo {title} {{Bayesian inference for compact binary coalescences with bilby: validation and application to the first LIGO\textendash{}Virgo gravitational-wave transient catalogue}},}\ }\href {\doibase 10.1093/mnras/staa2850} {\bibfield  {journal} {\bibinfo  {journal} {Mon. Not. Roy. Astron. Soc.}\ }\textbf {\bibinfo {volume} {499}},\ \bibinfo {pages} {3295--3319} (\bibinfo {year} {2020})},\ \Eprint {http://arxiv.org/abs/2006.00714} {arXiv:2006.00714 [astro-ph.IM]} \BibitemShut {NoStop}%
\bibitem [{\citenamefont {Smith}\ \emph {et~al.}(2020)\citenamefont {Smith}, \citenamefont {Ashton}, \citenamefont {Vajpeyi},\ and\ \citenamefont {Talbot}}]{Smith:2019ucc}%
  \BibitemOpen
  \bibfield  {author} {\bibinfo {author} {\bibfnamefont {Rory J.~E.}\ \bibnamefont {Smith}}, \bibinfo {author} {\bibfnamefont {Gregory}\ \bibnamefont {Ashton}}, \bibinfo {author} {\bibfnamefont {Avi}\ \bibnamefont {Vajpeyi}}, \ and\ \bibinfo {author} {\bibfnamefont {Colm}\ \bibnamefont {Talbot}},\ }\bibfield  {title} {\enquote {\bibinfo {title} {{Massively parallel Bayesian inference for transient gravitational-wave astronomy}},}\ }\href {\doibase 10.1093/mnras/staa2483} {\bibfield  {journal} {\bibinfo  {journal} {Mon. Not. Roy. Astron. Soc.}\ }\textbf {\bibinfo {volume} {498}},\ \bibinfo {pages} {4492--4502} (\bibinfo {year} {2020})},\ \Eprint {http://arxiv.org/abs/1909.11873} {arXiv:1909.11873 [gr-qc]} \BibitemShut {NoStop}%
\bibitem [{\citenamefont {Garc{\'\i}a-Quir{\'o}s}\ \emph {et~al.}(2020)\citenamefont {Garc{\'\i}a-Quir{\'o}s}, \citenamefont {Colleoni}, \citenamefont {Husa}, \citenamefont {Estell{\'e}s}, \citenamefont {Pratten}, \citenamefont {Ramos-Buades}, \citenamefont {Mateu-Lucena},\ and\ \citenamefont {Jaume}}]{Garcia-Quiros:2020qpx}%
  \BibitemOpen
  \bibfield  {author} {\bibinfo {author} {\bibfnamefont {Cecilio}\ \bibnamefont {Garc{\'\i}a-Quir{\'o}s}}, \bibinfo {author} {\bibfnamefont {Marta}\ \bibnamefont {Colleoni}}, \bibinfo {author} {\bibfnamefont {Sascha}\ \bibnamefont {Husa}}, \bibinfo {author} {\bibfnamefont {H{\'e}ctor}\ \bibnamefont {Estell{\'e}s}}, \bibinfo {author} {\bibfnamefont {Geraint}\ \bibnamefont {Pratten}}, \bibinfo {author} {\bibfnamefont {Antoni}\ \bibnamefont {Ramos-Buades}}, \bibinfo {author} {\bibfnamefont {Maite}\ \bibnamefont {Mateu-Lucena}}, \ and\ \bibinfo {author} {\bibfnamefont {Rafel}\ \bibnamefont {Jaume}},\ }\bibfield  {title} {\enquote {\bibinfo {title} {{Multimode frequency-domain model for the gravitational wave signal from nonprecessing black-hole binaries}},}\ }\href {\doibase 10.1103/PhysRevD.102.064002} {\bibfield  {journal} {\bibinfo  {journal} {Phys. Rev. D}\ }\textbf {\bibinfo {volume} {102}},\ \bibinfo {pages} {064002} (\bibinfo {year} {2020})},\ \Eprint {http://arxiv.org/abs/2001.10914} {arXiv:2001.10914 [gr-qc]} \BibitemShut {NoStop}%
\bibitem [{\citenamefont {Borghi}\ \emph {et~al.}(2024)\citenamefont {Borghi}, \citenamefont {Mancarella}, \citenamefont {Moresco}, \citenamefont {Tagliazucchi}, \citenamefont {Iacovelli}, \citenamefont {Cimatti},\ and\ \citenamefont {Maggiore}}]{Borghi:2023opd}%
  \BibitemOpen
  \bibfield  {author} {\bibinfo {author} {\bibfnamefont {Nicola}\ \bibnamefont {Borghi}}, \bibinfo {author} {\bibfnamefont {Michele}\ \bibnamefont {Mancarella}}, \bibinfo {author} {\bibfnamefont {Michele}\ \bibnamefont {Moresco}}, \bibinfo {author} {\bibfnamefont {Matteo}\ \bibnamefont {Tagliazucchi}}, \bibinfo {author} {\bibfnamefont {Francesco}\ \bibnamefont {Iacovelli}}, \bibinfo {author} {\bibfnamefont {Andrea}\ \bibnamefont {Cimatti}}, \ and\ \bibinfo {author} {\bibfnamefont {Michele}\ \bibnamefont {Maggiore}},\ }\bibfield  {title} {\enquote {\bibinfo {title} {{Cosmology and Astrophysics with Standard Sirens and Galaxy Catalogs in View of Future Gravitational Wave Observations}},}\ }\href {\doibase 10.3847/1538-4357/ad20eb} {\bibfield  {journal} {\bibinfo  {journal} {Astrophys. J.}\ }\textbf {\bibinfo {volume} {964}},\ \bibinfo {pages} {191} (\bibinfo {year} {2024})},\ \Eprint {http://arxiv.org/abs/2312.05302} {arXiv:2312.05302 [astro-ph.CO]} \BibitemShut {NoStop}%
\bibitem [{\citenamefont {Kiendrebeogo}\ \emph {et~al.}(2023)\citenamefont {Kiendrebeogo} \emph {et~al.}}]{Kiendrebeogo:2023hzf}%
  \BibitemOpen
  \bibfield  {author} {\bibinfo {author} {\bibfnamefont {R.~Weizmann}\ \bibnamefont {Kiendrebeogo}} \emph {et~al.},\ }\bibfield  {title} {\enquote {\bibinfo {title} {{Updated Observing Scenarios and Multimessenger Implications for the International Gravitational-wave Networks O4 and O5}},}\ }\href {\doibase 10.3847/1538-4357/acfcb1} {\bibfield  {journal} {\bibinfo  {journal} {Astrophys. J.}\ }\textbf {\bibinfo {volume} {958}},\ \bibinfo {pages} {158} (\bibinfo {year} {2023})},\ \Eprint {http://arxiv.org/abs/2306.09234} {arXiv:2306.09234 [astro-ph.HE]} \BibitemShut {NoStop}%
\bibitem [{\citenamefont {Mukherjee}(2022)}]{Mukherjee:2021rtw}%
  \BibitemOpen
  \bibfield  {author} {\bibinfo {author} {\bibfnamefont {Suvodip}\ \bibnamefont {Mukherjee}},\ }\bibfield  {title} {\enquote {\bibinfo {title} {{The redshift dependence of black hole mass distribution: is it reliable for standard sirens cosmology?}}}\ }\href {\doibase 10.1093/mnras/stac2152} {\bibfield  {journal} {\bibinfo  {journal} {Mon. Not. Roy. Astron. Soc.}\ }\textbf {\bibinfo {volume} {515}},\ \bibinfo {pages} {5495--5505} (\bibinfo {year} {2022})},\ \Eprint {http://arxiv.org/abs/2112.10256} {arXiv:2112.10256 [astro-ph.CO]} \BibitemShut {NoStop}%
\bibitem [{\citenamefont {Pierra}\ \emph {et~al.}(2024)\citenamefont {Pierra}, \citenamefont {Mastrogiovanni}, \citenamefont {Perri\`es},\ and\ \citenamefont {Mapelli}}]{Pierra:2023deu}%
  \BibitemOpen
  \bibfield  {author} {\bibinfo {author} {\bibfnamefont {Gr\'egoire}\ \bibnamefont {Pierra}}, \bibinfo {author} {\bibfnamefont {Simone}\ \bibnamefont {Mastrogiovanni}}, \bibinfo {author} {\bibfnamefont {St\'ephane}\ \bibnamefont {Perri\`es}}, \ and\ \bibinfo {author} {\bibfnamefont {Michela}\ \bibnamefont {Mapelli}},\ }\bibfield  {title} {\enquote {\bibinfo {title} {{Study of systematics on the cosmological inference of the Hubble constant from gravitational wave standard sirens}},}\ }\href {\doibase 10.1103/PhysRevD.109.083504} {\bibfield  {journal} {\bibinfo  {journal} {Phys. Rev. D}\ }\textbf {\bibinfo {volume} {109}},\ \bibinfo {pages} {083504} (\bibinfo {year} {2024})},\ \Eprint {http://arxiv.org/abs/2312.11627} {arXiv:2312.11627 [astro-ph.CO]} \BibitemShut {NoStop}%
\bibitem [{\citenamefont {Agarwal}\ \emph {et~al.}(2024)\citenamefont {Agarwal} \emph {et~al.}}]{Agarwal:2024hld}%
  \BibitemOpen
  \bibfield  {author} {\bibinfo {author} {\bibfnamefont {Aman}\ \bibnamefont {Agarwal}} \emph {et~al.},\ }\bibfield  {title} {\enquote {\bibinfo {title} {{Blinded Mock Data Challenge for Gravitational-Wave Cosmology-I: Assessing the Robustness of Methods Using Binary Black Holes Mass Spectrum}},}\ }\href@noop {} {\  (\bibinfo {year} {2024})},\ \Eprint {http://arxiv.org/abs/2412.14244} {arXiv:2412.14244 [astro-ph.CO]} \BibitemShut {NoStop}%
\bibitem [{\citenamefont {Karathanasis}\ \emph {et~al.}(2023)\citenamefont {Karathanasis}, \citenamefont {Mukherjee},\ and\ \citenamefont {Mastrogiovanni}}]{Karathanasis:2022rtr}%
  \BibitemOpen
  \bibfield  {author} {\bibinfo {author} {\bibfnamefont {Christos}\ \bibnamefont {Karathanasis}}, \bibinfo {author} {\bibfnamefont {Suvodip}\ \bibnamefont {Mukherjee}}, \ and\ \bibinfo {author} {\bibfnamefont {Simone}\ \bibnamefont {Mastrogiovanni}},\ }\bibfield  {title} {\enquote {\bibinfo {title} {{Binary black holes population and cosmology in new lights: signature of PISN mass and formation channel in GWTC-3}},}\ }\href {\doibase 10.1093/mnras/stad1373} {\bibfield  {journal} {\bibinfo  {journal} {Mon. Not. Roy. Astron. Soc.}\ }\textbf {\bibinfo {volume} {523}},\ \bibinfo {pages} {4539--4555} (\bibinfo {year} {2023})},\ \Eprint {http://arxiv.org/abs/2204.13495} {arXiv:2204.13495 [astro-ph.CO]} \BibitemShut {NoStop}%
\bibitem [{\citenamefont {Lalleman}\ \emph {et~al.}(2025)\citenamefont {Lalleman}, \citenamefont {Turbang}, \citenamefont {Callister},\ and\ \citenamefont {van Remortel}}]{Lalleman:2025xcs}%
  \BibitemOpen
  \bibfield  {author} {\bibinfo {author} {\bibfnamefont {Max}\ \bibnamefont {Lalleman}}, \bibinfo {author} {\bibfnamefont {Kevin}\ \bibnamefont {Turbang}}, \bibinfo {author} {\bibfnamefont {Thomas}\ \bibnamefont {Callister}}, \ and\ \bibinfo {author} {\bibfnamefont {Nick}\ \bibnamefont {van Remortel}},\ }\bibfield  {title} {\enquote {\bibinfo {title} {{No evidence that the binary black hole mass distribution evolves with redshift}},}\ }\href@noop {} {\  (\bibinfo {year} {2025})},\ \Eprint {http://arxiv.org/abs/2501.10295} {arXiv:2501.10295 [astro-ph.HE]} \BibitemShut {NoStop}%
\bibitem [{\citenamefont {Abbott}\ \emph {et~al.}(2021{\natexlab{d}})\citenamefont {Abbott} \emph {et~al.}}]{LIGOScientific:2019lzm}%
  \BibitemOpen
  \bibfield  {author} {\bibinfo {author} {\bibfnamefont {Rich}\ \bibnamefont {Abbott}} \emph {et~al.} (\bibinfo {collaboration} {LIGO Scientific, Virgo}),\ }\bibfield  {title} {\enquote {\bibinfo {title} {{Open data from the first and second observing runs of Advanced LIGO and Advanced Virgo}},}\ }\href {\doibase 10.1016/j.softx.2021.100658} {\bibfield  {journal} {\bibinfo  {journal} {SoftwareX}\ }\textbf {\bibinfo {volume} {13}},\ \bibinfo {pages} {100658} (\bibinfo {year} {2021}{\natexlab{d}})},\ \Eprint {http://arxiv.org/abs/1912.11716} {arXiv:1912.11716 [gr-qc]} \BibitemShut {NoStop}%
\bibitem [{\citenamefont {Abbott}\ \emph {et~al.}(2023{\natexlab{d}})\citenamefont {Abbott} \emph {et~al.}}]{KAGRA:2023pio}%
  \BibitemOpen
  \bibfield  {author} {\bibinfo {author} {\bibfnamefont {R.}~\bibnamefont {Abbott}} \emph {et~al.} (\bibinfo {collaboration} {KAGRA, VIRGO, LIGO Scientific}),\ }\bibfield  {title} {\enquote {\bibinfo {title} {{Open Data from the Third Observing Run of LIGO, Virgo, KAGRA, and GEO}},}\ }\href {\doibase 10.3847/1538-4365/acdc9f} {\bibfield  {journal} {\bibinfo  {journal} {Astrophys. J. Suppl.}\ }\textbf {\bibinfo {volume} {267}},\ \bibinfo {pages} {29} (\bibinfo {year} {2023}{\natexlab{d}})},\ \Eprint {http://arxiv.org/abs/2302.03676} {arXiv:2302.03676 [gr-qc]} \BibitemShut {NoStop}%
\bibitem [{\citenamefont {Speagle}(2020)}]{Speagle:2019ivv}%
  \BibitemOpen
  \bibfield  {author} {\bibinfo {author} {\bibfnamefont {Joshua~S.}\ \bibnamefont {Speagle}},\ }\bibfield  {title} {\enquote {\bibinfo {title} {{dynesty: a dynamic nested sampling package for estimating Bayesian posteriors and evidences}},}\ }\href {\doibase 10.1093/mnras/staa278} {\bibfield  {journal} {\bibinfo  {journal} {Mon. Not. Roy. Astron. Soc.}\ }\textbf {\bibinfo {volume} {493}},\ \bibinfo {pages} {3132--3158} (\bibinfo {year} {2020})},\ \Eprint {http://arxiv.org/abs/1904.02180} {arXiv:1904.02180 [astro-ph.IM]} \BibitemShut {NoStop}%
\bibitem [{\citenamefont {Koposov}\ \emph {et~al.}(2024)\citenamefont {Koposov} \emph {et~al.}}]{dynesty}%
  \BibitemOpen
  \bibfield  {author} {\bibinfo {author} {\bibfnamefont {Sergey}\ \bibnamefont {Koposov}} \emph {et~al.},\ }\href {\doibase 10.5281/zenodo.12537467} {\enquote {\bibinfo {title} {dynesty},}\ } (\bibinfo {year} {2024})\BibitemShut {NoStop}%
\bibitem [{\citenamefont {Skilling}(2006)}]{Skilling2006NestedSF}%
  \BibitemOpen
  \bibfield  {author} {\bibinfo {author} {\bibfnamefont {John}\ \bibnamefont {Skilling}},\ }\bibfield  {title} {\enquote {\bibinfo {title} {Nested sampling for general bayesian computation},}\ }\href {https://api.semanticscholar.org/CorpusID:652102} {\bibfield  {journal} {\bibinfo  {journal} {Bayesian Analysis}\ }\textbf {\bibinfo {volume} {1}},\ \bibinfo {pages} {833--859} (\bibinfo {year} {2006})}\BibitemShut {NoStop}%
\bibitem [{\citenamefont {Williams}(2021)}]{nessai}%
  \BibitemOpen
  \bibfield  {author} {\bibinfo {author} {\bibfnamefont {Michael~J.}\ \bibnamefont {Williams}},\ }\href {\doibase 10.5281/zenodo.4550693} {\enquote {\bibinfo {title} {nessai: Nested sampling with artificial intelligence},}\ } (\bibinfo {year} {2021})\BibitemShut {NoStop}%
\bibitem [{\citenamefont {Williams}\ \emph {et~al.}(2021)\citenamefont {Williams}, \citenamefont {Veitch},\ and\ \citenamefont {Messenger}}]{Williams:2021qyt}%
  \BibitemOpen
  \bibfield  {author} {\bibinfo {author} {\bibfnamefont {Michael~J.}\ \bibnamefont {Williams}}, \bibinfo {author} {\bibfnamefont {John}\ \bibnamefont {Veitch}}, \ and\ \bibinfo {author} {\bibfnamefont {Chris}\ \bibnamefont {Messenger}},\ }\bibfield  {title} {\enquote {\bibinfo {title} {{Nested sampling with normalizing flows for gravitational-wave inference}},}\ }\href {\doibase 10.1103/PhysRevD.103.103006} {\bibfield  {journal} {\bibinfo  {journal} {Phys. Rev. D}\ }\textbf {\bibinfo {volume} {103}},\ \bibinfo {pages} {103006} (\bibinfo {year} {2021})},\ \Eprint {http://arxiv.org/abs/2102.11056} {arXiv:2102.11056 [gr-qc]} \BibitemShut {NoStop}%
\bibitem [{\citenamefont {Williams}\ \emph {et~al.}(2023)\citenamefont {Williams}, \citenamefont {Veitch},\ and\ \citenamefont {Messenger}}]{Williams:2023ppp}%
  \BibitemOpen
  \bibfield  {author} {\bibinfo {author} {\bibfnamefont {Michael~J.}\ \bibnamefont {Williams}}, \bibinfo {author} {\bibfnamefont {John}\ \bibnamefont {Veitch}}, \ and\ \bibinfo {author} {\bibfnamefont {Chris}\ \bibnamefont {Messenger}},\ }\bibfield  {title} {\enquote {\bibinfo {title} {{Importance nested sampling with normalising flows}},}\ }\href {\doibase 10.1088/2632-2153/acd5aa} {\bibfield  {journal} {\bibinfo  {journal} {Mach. Learn. Sci. Tech.}\ }\textbf {\bibinfo {volume} {4}},\ \bibinfo {pages} {035011} (\bibinfo {year} {2023})},\ \Eprint {http://arxiv.org/abs/2302.08526} {arXiv:2302.08526 [astro-ph.IM]} \BibitemShut {NoStop}%
\bibitem [{\citenamefont {Lange}\ \emph {et~al.}(2018)\citenamefont {Lange}, \citenamefont {O'Shaughnessy},\ and\ \citenamefont {Rizzo}}]{Lange:2018pyp}%
  \BibitemOpen
  \bibfield  {author} {\bibinfo {author} {\bibfnamefont {Jacob}\ \bibnamefont {Lange}}, \bibinfo {author} {\bibfnamefont {Richard}\ \bibnamefont {O'Shaughnessy}}, \ and\ \bibinfo {author} {\bibfnamefont {Monica}\ \bibnamefont {Rizzo}},\ }\bibfield  {title} {\enquote {\bibinfo {title} {{Rapid and accurate parameter inference for coalescing, precessing compact binaries}},}\ }\href@noop {} {\  (\bibinfo {year} {2018})},\ \Eprint {http://arxiv.org/abs/1805.10457} {arXiv:1805.10457 [gr-qc]} \BibitemShut {NoStop}%
\bibitem [{\citenamefont {Gaebel}\ \emph {et~al.}(2019)\citenamefont {Gaebel}, \citenamefont {Veitch}, \citenamefont {Dent},\ and\ \citenamefont {Farr}}]{Gaebel:2018poe}%
  \BibitemOpen
  \bibfield  {author} {\bibinfo {author} {\bibfnamefont {Sebastian~M.}\ \bibnamefont {Gaebel}}, \bibinfo {author} {\bibfnamefont {John}\ \bibnamefont {Veitch}}, \bibinfo {author} {\bibfnamefont {Thomas}\ \bibnamefont {Dent}}, \ and\ \bibinfo {author} {\bibfnamefont {Will~M.}\ \bibnamefont {Farr}},\ }\bibfield  {title} {\enquote {\bibinfo {title} {{Digging the population of compact binary mergers out of the noise}},}\ }\href {\doibase 10.1093/mnras/stz225} {\bibfield  {journal} {\bibinfo  {journal} {Mon. Not. Roy. Astron. Soc.}\ }\textbf {\bibinfo {volume} {484}},\ \bibinfo {pages} {4008--4023} (\bibinfo {year} {2019})},\ \Eprint {http://arxiv.org/abs/1809.03815} {arXiv:1809.03815 [astro-ph.IM]} \BibitemShut {NoStop}%
\bibitem [{\citenamefont {Mandel}\ \emph {et~al.}(2020)\citenamefont {Mandel}, \citenamefont {M\"uller}, \citenamefont {Riley}, \citenamefont {de~Mink}, \citenamefont {Vigna-G\'omez},\ and\ \citenamefont {Chattopadhyay}}]{Mandel:2020cig}%
  \BibitemOpen
  \bibfield  {author} {\bibinfo {author} {\bibfnamefont {Ilya}\ \bibnamefont {Mandel}}, \bibinfo {author} {\bibfnamefont {Bernhard}\ \bibnamefont {M\"uller}}, \bibinfo {author} {\bibfnamefont {Jeff}\ \bibnamefont {Riley}}, \bibinfo {author} {\bibfnamefont {Selma~E.}\ \bibnamefont {de~Mink}}, \bibinfo {author} {\bibfnamefont {Alejandro}\ \bibnamefont {Vigna-G\'omez}}, \ and\ \bibinfo {author} {\bibfnamefont {Debatri}\ \bibnamefont {Chattopadhyay}},\ }\bibfield  {title} {\enquote {\bibinfo {title} {{Binary population synthesis with probabilistic remnant mass and kick prescriptions}},}\ }\href {\doibase 10.1093/mnras/staa3390} {\bibfield  {journal} {\bibinfo  {journal} {Mon. Not. Roy. Astron. Soc.}\ }\textbf {\bibinfo {volume} {500}},\ \bibinfo {pages} {1380--1384} (\bibinfo {year} {2020})},\ \Eprint {http://arxiv.org/abs/2007.03890} {arXiv:2007.03890 [astro-ph.HE]} \BibitemShut {NoStop}%
\bibitem [{\citenamefont {Kapadia}\ \emph {et~al.}(2020)\citenamefont {Kapadia} \emph {et~al.}}]{Kapadia:2019uut}%
  \BibitemOpen
  \bibfield  {author} {\bibinfo {author} {\bibfnamefont {Shasvath~J.}\ \bibnamefont {Kapadia}} \emph {et~al.},\ }\bibfield  {title} {\enquote {\bibinfo {title} {{A self-consistent method to estimate the rate of compact binary coalescences with a Poisson mixture model}},}\ }\href {\doibase 10.1088/1361-6382/ab5f2d} {\bibfield  {journal} {\bibinfo  {journal} {Class. Quant. Grav.}\ }\textbf {\bibinfo {volume} {37}},\ \bibinfo {pages} {045007} (\bibinfo {year} {2020})},\ \Eprint {http://arxiv.org/abs/1903.06881} {arXiv:1903.06881 [astro-ph.HE]} \BibitemShut {NoStop}%
\bibitem [{\citenamefont {Vitale}\ \emph {et~al.}(2020)\citenamefont {Vitale}, \citenamefont {Gerosa}, \citenamefont {Farr},\ and\ \citenamefont {Taylor}}]{Vitale:2020aaz}%
  \BibitemOpen
  \bibfield  {author} {\bibinfo {author} {\bibfnamefont {Salvatore}\ \bibnamefont {Vitale}}, \bibinfo {author} {\bibfnamefont {Davide}\ \bibnamefont {Gerosa}}, \bibinfo {author} {\bibfnamefont {Will~M.}\ \bibnamefont {Farr}}, \ and\ \bibinfo {author} {\bibfnamefont {Stephen~R.}\ \bibnamefont {Taylor}},\ }\bibfield  {title} {\enquote {\bibinfo {title} {{Inferring the properties of a population of compact binaries in presence of selection effects}},}\ }\href {\doibase 10.1007/978-981-15-4702-7-45-1} {\  (\bibinfo {year} {2020}),\ 10.1007/978-981-15-4702-7-45-1},\ \Eprint {http://arxiv.org/abs/2007.05579} {arXiv:2007.05579 [astro-ph.IM]} \BibitemShut {NoStop}%
\bibitem [{\citenamefont {Essick}\ and\ \citenamefont {Fishbach}(2024)}]{Essick:2023upv}%
  \BibitemOpen
  \bibfield  {author} {\bibinfo {author} {\bibfnamefont {Reed}\ \bibnamefont {Essick}}\ and\ \bibinfo {author} {\bibfnamefont {Maya}\ \bibnamefont {Fishbach}},\ }\bibfield  {title} {\enquote {\bibinfo {title} {{Ensuring Consistency between Noise and Detection in Hierarchical Bayesian Inference}},}\ }\href {\doibase 10.3847/1538-4357/ad1604} {\bibfield  {journal} {\bibinfo  {journal} {Astrophys. J.}\ }\textbf {\bibinfo {volume} {962}},\ \bibinfo {pages} {169} (\bibinfo {year} {2024})},\ \Eprint {http://arxiv.org/abs/2310.02017} {arXiv:2310.02017 [gr-qc]} \BibitemShut {NoStop}%
\bibitem [{\citenamefont {Collaboration}\ \emph {et~al.}(2023)\citenamefont {Collaboration}, \citenamefont {Collaboration},\ and\ \citenamefont {Collaboration}}]{O3:bbhpop}%
  \BibitemOpen
  \bibfield  {author} {\bibinfo {author} {\bibfnamefont {LIGO~Scientific}\ \bibnamefont {Collaboration}}, \bibinfo {author} {\bibfnamefont {Virgo}\ \bibnamefont {Collaboration}}, \ and\ \bibinfo {author} {\bibfnamefont {KAGRA}\ \bibnamefont {Collaboration}},\ }\href@noop {} {\enquote {\bibinfo {title} {Gwtc-3: Compact binary coalescences observed by ligo and virgo during the second part of the third observing run — o3 search sensitivity estimates [data set]},}\ }\bibinfo {howpublished} {\url{https://doi.org/10.5281/zenodo.7890437}} (\bibinfo {year} {2023})\BibitemShut {NoStop}%
\bibitem [{\citenamefont {Talbot}\ and\ \citenamefont {Golomb}(2023)}]{Talbot:2023pex}%
  \BibitemOpen
  \bibfield  {author} {\bibinfo {author} {\bibfnamefont {Colm}\ \bibnamefont {Talbot}}\ and\ \bibinfo {author} {\bibfnamefont {Jacob}\ \bibnamefont {Golomb}},\ }\bibfield  {title} {\enquote {\bibinfo {title} {{Growing pains: understanding the impact of likelihood uncertainty on hierarchical Bayesian inference for gravitational-wave astronomy}},}\ }\href {\doibase 10.1093/mnras/stad2968} {\bibfield  {journal} {\bibinfo  {journal} {Mon. Not. Roy. Astron. Soc.}\ }\textbf {\bibinfo {volume} {526}},\ \bibinfo {pages} {3495--3503} (\bibinfo {year} {2023})},\ \Eprint {http://arxiv.org/abs/2304.06138} {arXiv:2304.06138 [astro-ph.IM]} \BibitemShut {NoStop}%
\end{thebibliography}%

\end{document}